\documentclass[prd,aps,twocolumn,preprintnumbers,amsmath,amssymb,nofootinbib]{revtex4}
\usepackage{amsmath,amssymb} 

\usepackage{graphicx}

\newcommand{\amp}{&\!\!}
\newcommand{\gmo}{\psi}
\newcommand{\mo}{\psi}
\newcommand{\gmog}{b}
\newcommand{\gmoa}{a}
\newcommand{\gmoi}{\psi_i}
\newcommand{\gmogi}{b_i}
\newcommand{\gmoai}{a_i}

\newcommand{\mokg}{b}
\newcommand{\mokgi}{b_i}
\newcommand{\moka}{a}
\newcommand{\mokai}{a_i}
\newcommand{\moag}{b_4}
\newcommand{\moaa}{a_4}
\newcommand{\mokgo}{b_1}
\newcommand{\mokgt}{b_2}
\newcommand{\mokgth}{b_3}
\newcommand{\mokao}{a_1}
\newcommand{\mokat}{a_2}
\newcommand{\mokath}{a_3}

\newcommand{\moki}{\psi_i}
\newcommand{\moko}{\psi_1}
\newcommand{\mokt}{\psi_2}
\newcommand{\mokth}{\psi_3}
\newcommand{\moa}{\psi_4}

\newcommand{\fz}{f_0}
\newcommand{\fth}{f_3}
\newcommand{\ffi}{f_{4,i}}
\newcommand{\ffo}{f_{4,1}}
\newcommand{\fft}{f_{4,2}}
\newcommand{\ffth}{f_{4,3}}
\newcommand{\fs}{f_6}
\newcommand{\hfz}{\hat{f}_0}
\newcommand{\hfth}{\hat{f}_3}
\newcommand{\hffi}{\hat{f}_{4,i}}
\newcommand{\hffo}{\hat{f}_{4,1}}
\newcommand{\hfft}{\hat{f}_{4,2}}
\newcommand{\hffth}{\hat{f}_{4,3}}

\newcommand{\sgn}{\delta}

\newcommand{\vmoki}{\psi_i}
\newcommand{\vmoko}{\psi_1}
\newcommand{\vmokt}{\psi_2}
\newcommand{\vmokth}{\psi_3}
\newcommand{\vmoa}{\psi_4}
\newcommand{\vmoco}{\psi_5}
\newcommand{\vmoct}{\psi_6}
\newcommand{\vmocth}{\psi_7}

\newcommand{\vmokgi}{b_i}

\newcommand{\vmokai}{a_i}
\newcommand{\vmokgo}{b_1}
\newcommand{\vmokgt}{b_2}
\newcommand{\vmokgth}{b_3}
\newcommand{\vmokao}{a_1}
\newcommand{\vmokat}{a_2}
\newcommand{\vmokath}{a_3}
\newcommand{\vmoag}{b_4}
\newcommand{\vmoaa}{a_4}
\newcommand{\vmoaah}{\hat{a}_4}
\newcommand{\vmoci}{\psi_i}
\newcommand{\vmocai}{a_i}
\newcommand{\vmocgi}{b_i}
\newcommand{\vmocgo}{b_5}
\newcommand{\vmocgt}{b_6}
\newcommand{\vmocgth}{b_7}
\newcommand{\vmocao}{a_5}
\newcommand{\vmocat}{a_6}
\newcommand{\vmocath}{a_7}
\newcommand{\lamooo}{f_{111}}
\newcommand{\lamooop}{f_{111}'}
\newcommand{\lamoot}{f_{112}}
\newcommand{\lamootp}{f_{112}'}
\newcommand{\lamoof}{f_{114}}
\newcommand{\lamooth}{f_{113}}
\newcommand{\lamott}{f_{122}}
\newcommand{\lamotf}{f_{124}}
\newcommand{\lamttt}{f_{222}}

\newcommand{\imoki}{\psi_i}

\newcommand{\imokgi}{b_i}
\newcommand{\imokgj}{b_j}

\newcommand{\imokai}{a_i}
\newcommand{\imokaj}{a_j}
\newcommand{\imoag}{b_5}
\newcommand{\imoaa}{a_5}
\newcommand{\imokgo}{b_1}
\newcommand{\imokgt}{b_2}
\newcommand{\imokgth}{b_3}
\newcommand{\imokgf}{b_4}
\newcommand{\imokao}{a_1}

\newcommand{\imoko}{\psi_1}
\newcommand{\imokt}{\psi_2}
\newcommand{\imokth}{\psi_3}
\newcommand{\imokf}{\psi_4}
\newcommand{\imoa}{\psi_5}

\newcommand{\fff}{f_{4,4}}
\newcommand{\hfff}{\hat{f}_{4,4}}
\newcommand{\imoc}{\psi_6}
\newcommand{\imocg}{b_6}
\newcommand{\imoca}{a_6}
\newcommand{\sgnot}{\delta_{12}}
\newcommand{\sgnthz}{\delta_{30}}

\newcommand{\None}{p_1}
\newcommand{\Ntwo}{p_2}
\newcommand{\tO}{q_1}
\newcommand{\sO}{q_2}
\newcommand{\uO}{q_3}

\newcommand{\rat}{s}
\newcommand{\flx}{t}

\newcommand{\mpl}{\bar{m}_{\mbox{\tiny{P}}}}
\newcommand{\VE}{\bar{V}}

\def\ie{{\frenchspacing\it i.e.}}
\def\eg{{\frenchspacing\it e.g.}}
\def\etc{{\frenchspacing\it etc.}}

\def\GeV{{\rm GeV}}

\def\beq#1{\begin{equation}\label{#1}}
\def\eeq{\end{equation}}
\def\beqa#1{\begin{eqnarray}\label{#1}}
\def\eeqa{\end{eqnarray}}

\def\ns{{n_s}}
\def\nt{{n_t}}
\def\al{\alpha}

\def\rhoend{\rho_{\rm end}}

\def\phiend{{\phi_{\rm end}}}

\def\rhoend{{\rho_{\rm end}}}
\def\rhoreh{{\rho_{\rm reh}}}

\def\mbar{{\bar m}}

\def\spose#1{\hbox to 0pt{#1\hss}}
\def\simlt{\mathrel{\spose{\lower 3pt\hbox{$\mathchar"218$}}
     \raise 2.0pt\hbox{$\mathchar"13C$}}}
\def\simgt{\mathrel{\spose{\lower 3pt\hbox{$\mathchar"218$}}
     \raise 2.0pt\hbox{$\mathchar"13E$}}}
\def\simpropto{\mathrel{\spose{\lower 3pt\hbox{$\mathchar"218$}}
     \raise 2.0pt\hbox{$\propto$}}}

\begin{document}

\title{
Searching for Inflation in Simple String Theory Models: An Astrophysical Perspective\\
}

\author{
Mark P. Hertzberg$^1$\footnote{Electronic address: mphertz@mit.edu},
 Max Tegmark$^1$, Shamit Kachru$^2$, Jessie Shelton$^{1,3}$, and Onur \"Ozcan$^1$} 
\address{$^1$Dept.~of Physics, Massachusetts Institute of Technology, Cambridge, MA 02139, USA}
\address{$^2$Dept.~of Physics and SLAC, Stanford University, Stanford, CA 94305, USA}
\address{$^3$Dept.~of Physics and Astronomy, Rutgers University, Piscataway, NJ 08855, USA}


\begin{abstract}
Attempts to connect string theory with astrophysical observation are hampered by a 
jargon barrier, where an intimidating profusion of orientifolds, K\"ahler potentials, {\etc}
dissuades cosmologists from attempting to 
work out the astrophysical observables of specific string theory solutions from the recent literature.
We attempt to help bridge this gap by giving a pedagogical exposition with detailed 
examples, aimed at astrophysicists and high energy theorists alike, of how to compute
predictions for familiar cosmological parameters when starting with a 10-dimensional string theory action.
This is done by investigating inflation in string theory, since inflation is the dominant paradigm 
for how early universe physics determines cosmological parameters.


We analyze three explicit string models from the recent literature, each containing
an infinite number of ``vacuum" solutions.
Our numerical investigation of some natural candidate inflatons, the so-called ``moduli fields,''
fails to find inflation.
We also find in the simplest models that, after suitable field redefinitions, vast numbers of these vacua differ only in an overall 
constant multiplying the effective inflaton potential, a difference which affects neither the 
potential's shape nor its ability to support slow-roll inflation.
This illustrates that even having an infinite number of vacua does not guarantee having inflating ones.
This may be an artifact of the simplicity of the models that we study.
Instead, more complicated string theory models appear to be required, suggesting that
identifying the inflating subset of the string landscape will be challenging.

\end{abstract}

\vspace*{-\bigskipamount} \preprint{MIT-CTP-3859}\preprint{SU-ITP-07/13}\preprint{SLAC-PUB-12778}

\maketitle

\section{Introduction}

String theory is currently the most popular candidate for a consistent theory of quantum gravity, 
but the goal of confronting it with observation remains elusive.
It is sometimes said that testing string theory requires prohibitively high energy accelerators,
in order to probe the Planck scale predictions of the theory.
There are non-trivial tests of string theory at low energies, however, such as the requirement
to have a solution with the standard model of particle physics.  
Furthermore, it is plausible that  we can test string theory by turning to cosmology.
The earliest moments of our universe involved extreme energies and the fingerprints
of its birth are revealed today by
precision measurements of the cosmic microwave background \cite{CMB} and the large-scale structure of the universe \cite{LSS,Sanchez06}. A highly non-trivial test of string theory then is whether it can
reproduce our cosmology.
With inflation emerging as the paradigm of early universe phenomena,
string theory or any competing theory of quantum gravity must be able to realize this.
Moreover, merely producing many $e$-foldings of inflation is not good enough: the details of inflation must give correct predictions for as many as eight cosmological parameters which have been 
measured or constrained \cite{inflation_astro-ph/0410281}.

Although there have been substantial efforts in the string theory literature aimed at identifying and counting 
long-lived potential energy minima (so-called vacua)
\cite{evareview,Grana,Douglas,Lust_1,Lust_2,Lust_3}, 
the key cosmological observables depend also on the
history of how our spacetime region evolved to this minimum by slow-roll inflation and/or tunneling.
This article is aimed at discussing one of the simplest realizations of slow-roll inflation from string theory,
where the inflaton fields are the 
so-called moduli which, loosely speaking, correspond to the size and shape of curled up extra 
dimensions (see Table \ref{deftable}) \cite{modinf_1,modinf_2}.
An alternative scenario involves 
dynamical branes \cite{Tye_1,Tye_2,Tye_3,Tye_4,Panda,Bean,KKLMMT}, with the most explicit models to date appearing in \cite{Baumann_1,Baumann_2}.  Some other possibilities include \cite{Alexander, Beatriz}. For recent reviews of inflation in string theory, see \cite{infrev_1,infrev_2,infrev_3,infrev_4}.

This paper is aimed at anyone who is intrigued by the possibility of connecting string theory
and cosmology. We hope that it is accessible to non-string theorists,
and have therefore tried hard to minimize string theory specific terminology and notation, 
referring the interested reader to more technical references for further detail.
Table \ref{deftable} provides a ``Stringlish to English" reference dictionary for the most central string theory terms.
Bridging the gap between string theory and observational astrophysics is important for both fields:
not only does it offer potential tests of string theory as mentioned above, but it also 
offers an opportunity for cosmologists to move beyond the tradition of 
putting in inflaton potentials by hand.

\begin{table*}
\caption{Dictionary of some basic string theory terminology.}
\label{deftable}
\begin{tabular*}{16.9cm}[t]{ c | l | p{10.4cm} }
\hline
Symbol & Name & Approximate meaning\\
\hline
$\alpha'$ & Regge parameter & Inverse string tension \\
$l_s$ & String length & $= 2\pi\sqrt{\alpha'}$ (in our convention) \\
$\kappa_{10}$ & 10-d gravitational strength & $=\sqrt{8\pi G_{10}}=l_s^4/\sqrt{4\pi}$, gravitational strength in 10 dims\\
$\mpl$ & (Reduced) Planck mass & $= 1/\sqrt{8\pi G}$, mass scale of quantum gravity in 4 dims\\ 
$\phi$ & Dilaton & Scalar field that rescales the strength of gravity \\
$a_i$ & Axions & Pseudo-scalars that appear in the 4-d theory \\
$b_i$ & Geometric moduli & Scalar fields describing $\phi$ and the size \& shape of the compact space \\
           & \,\,-- Dilaton modulus\footnote{When partnered with its axion, the dilaton modulus is known as the ``axio-dilaton".} & $\sim e^{-\phi}$ (explicit form is model dependent) \\
           & \,\,-- K\"ahler moduli & Scalar fields that specify the {\em size} of the compact space\\
           & \,\,-- Complex structure moduli & Scalar fields that specify the {\em shape} of the compact space\\
$\psi_i$ & Complex moduli & $= a_i + i\,b_i$\\
$\boldsymbol{\psi}$ & Complex inflaton vector & $=(\psi_1,...,\psi_n)$, the complex moduli-vector that can evolve during inflation \\
$\boldsymbol{\phi}$ & Real inflaton vector & $=(a_1,b_1,...,a_n,b_n)$, the real moduli-vector that can evolve during inflation \\
$g_s$ & String coupling & $= e^{\phi}$, the string loop expansion parameter \\
$F_p$ & $p$-form field strength & Generalized electromagnetic field strength carrying $p$-indices \\
$f_p$ & Flux & $\propto\int F_p$, (normally integer valued)
equivalent to a generalized electric or magnetic charge, 
but can arise purely due to non-trivial topology \\
$g_{10}$/$R_{10}$ & 10-d string metric/Ricci scalar & Metric/Ricci scalar in the fundamental 10-d action in string frame \\
$g_{4}$/$R_{4}$ & 4-d string metric/Ricci scalar & Metric/Ricci scalar in the effective 4-d action in string frame \\
$g_{E}$/$R_{E}$ & 4-d Einstein metric/Ricci scalar & Metric/Ricci scalar in the effective 4-d action after a conformal transformation to Einstein frame\\
$g_6$ & Metric on compact space & 2nd block of $g_{10}=\mbox{diag}(g_4,g_6)$, describing
the geometry of compact space \\
Vol & 6-d volume of compact space & $= \int_{\mbox{\tiny{cs}}} d^6 x\sqrt{g_6}$ 
(cs $\equiv$ compact space) \\
$T^6$ & 6-d torus & A 6-d manifold that is Riemann flat, defined by periodic identifications \\
$K$ & K\"ahler potential & Scalar function whose Hessian matrix is the metric on moduli space \\
$W$ & Superpotential & Scalar function that describes the interactions between moduli set up by fluxes etc in a supersymmetric theory in 4 dims \\
$V$ & Supergravity potential energy & Potential energy function governing the fields $a_i,\,b_i$ 
in 4 dims as set up by the supergravity formula in eq.~(\ref{SUGRA}) \\
$\VE$ & Potential energy & $=\mpl^4\,V/4\pi$, potential energy function in 4 dims in conventional units \\
$\epsilon$ & First slow-roll parameter & See eq.~(\ref{SReps}), quantifies the magnitude of the 1st derivatives of $V$ \\
$\eta$ & Second slow-roll parameter & See eq.~(\ref{SReta}), quantifies the minimum of the 2nd derivatives of $V$ \\
- & D(irichlet)$p$-brane & A $(p+1)$-d object that contributes positive energy and can 
source $F_{p+2}$ \\
- & O(rientifold)$p$-plane & A $(p+1)$-d plane that can contribute negative energy; it arises at fixed points in so-called orientifold models \\
\hline
\end{tabular*}
\end{table*}

\subsection{Can string theory describe inflation?}

Answering the question of whether inflation can be embedded in string theory
is very difficult.
First of all, there is no known complete formulation of string theory, so the theory is 
not fully understood.
In particular there does not seem to be a dynamical mechanism which selects
the way in which the theory, which lives in 10 dimensions, should be compactified to 4 dimensions.
Instead, there is apparently a ``landscape'' of possible 4-dimensional effective physics theories.
Each so-called vacuum in the landscape corresponds to a stable or very long-lived configuration,
containing, amongst other things, gravity, scalar fields 
(the above-mentioned moduli), 
and various potential energies.
 We would like to know if some of these vacua reproduce our observed large and rather uniform patch of 3+1-dimensional spacetime.
We return below to the process by which this 4-dimensional picture emerges from the
10-dimensional picture. What is exciting is that these ingredients in 4 dimensions are 
precisely those used in inflationary model building.

There are several difficulties that must be overcome to build reasonable 4-dimensional models.
Firstly, it is difficult to stabilize the moduli.
One reason this is problematic is that one of the moduli corresponds to the size of the compact space. If it were not stabilized then the field may roll to very large values and the ``compact space" would de-compactify. Furthermore, stabilizing all moduli is important to reproduce our universe which exhibits
(approximate) Poincar\'e invariance, and a notable absence of long-range fifth forces.
Authors have discussed various ingredients that may be included for such stabilization, \eg,
non-perturbative effects that go by names such as gluino condensation and instantons.
A second difficulty arises due to supersymmetry; most well-understood vacua have negative cosmological constants, \ie, correspond to anti de Sitter spaces
(we will discuss this issue in some detail later on).
One resolution of these problems was 
provided by KKLT \cite{KKLT}, who included non-perturbative phenomena for stabilization and 
broke supersymmetry to achieve a 4-dimensional solution with positive cosmological constant.  (Earlier
constructions of de Sitter vacua in non-critical string theory appeared in \cite{Eva}).

Within this framework (where non-perturbative corrections play an important role, and
a supersymmetry breaking sector is incorporated to generate positive vacuum
energy) 
various plausible models of inflation using moduli fields have been suggested 
in the literature: ``N-flation" \cite{Nflation_1,Nflation_2}, ``K\"ahler moduli
inflation" \cite{Conlon1_1,Conlon1_2,Conlon1_3},  ``Inflating in a Better Racetrack" \cite{Racetrack},
and using brane moduli ``KKLMMT" and related scenarios
\cite{KKLMMT,Baumann_1,Baumann_2,DBI}.
These models, however, share a common property: they are not entirely explicit constructions,
though steady progress in that direction has been made.
This leads us to the obvious and important question: {\it Can we realize inflation explicitly and reliably
in string theory?}

\subsection{Explicit string theory inflation}

One of the difficulties with making fully explicit models of inflation has been that most of
the methods of moduli stabilization involve an interplay of classical effects in the potential
(which are easily computable), and quantum effects whose existence is well established, but for
which precise computations are often difficult.

However, recently, models which stabilize all moduli using classical effects alone have been
constructed.
These manage to stabilize all moduli 
in a regime where all approximations are parametrically under
control \cite{Grimm,Derendinger,DeWolfe, Villadoro, Ihl}. 
These examples are all explicit stable compactifications. 
They primarily achieve this stability by using potential energy contributions from generalized 
electric and magnetic fields (so-called fluxes; see Table \ref{deftable}) 
whose combined energy are minimized when the moduli fields take some particular values.
To borrow the language of quantum field theory, the potential functions in these models are
generated at ``tree-level", and quantum corrections are shown to be small.
This stabilization of all moduli at tree-level is what distinguishes these models from their
earlier counterparts.\footnote{It should be noted that the parametric control (which arises
at very large values of fluxes) comes with various features which are undesirable for phenomenology:
the extra dimensions become large at large flux values, the moduli masses become small,
and the coupling constants become extremely weak.  So for real world model-building, one
would place a cutoff on the flux values, and lose parametric control.
However, as simple and explicit examples
of stable compactifications, these examples provide a useful setting to address theoretical questions,
like our question about explicit computable models of moduli inflation.}

In this paper, we take three such recently found models and analyze each from the
point of view of inflation.
Specifically, we consider the models of 
DeWolfe, Giryavets, Kachru, \& Taylor (DGKT) \cite{DeWolfe},
Villadoro \& Zwirner (VZ) \cite{Villadoro}, and Ihl \& Wrase (IW) \cite{Ihl}.
All of these arise in the string theory known as type IIA. 
Each of these models possesses an infinite number of vacua, distinguished by fluxes.

We wish to examine whether the tree-level potential for moduli fields in these models can
support inflation.
A well-known challenge for generating inflation within string theory is that generic 
potentials will not be sufficiently flat, a point we will expand on below.
However, one might hope that when vast or infinite numbers of vacua are available, some of them would by chance 
have sufficiently flat directions to support inflationary slow-roll even if generic ones do not.\footnote{This of course depends on the extent to which the 
inflationary slow-roll parameters $\epsilon$ and $\eta$ vary as the fluxes are changed: if they densely sample a wide range including 
$\epsilon<1$, $|\eta|<1$, then flux tuning can allow inflation, otherwise even  large numbers of vacua may not help.}
One of our key findings 
below is that in the case of the (simplest moduli in the simplest examples of) IIA flux vacua 
that we study here, the
distributions of the quantities relevant for inflation are narrow enough that the
large number of vacua does not help.  Instead, the candidate inflaton potentials have
the same shape in many (sometimes all) of the vacua, differing only in overall normalization.
So, somewhat surprisingly, our search below does not turn up a single vacuum supporting inflation.

We hasten to emphasize that since 
these models are in many ways the simplest possible models in their class (involving the
simplest compactification geometry, the six-torus, in a crucial way), and since we focus
on a simple subset of the moduli (the ``untwisted moduli") even in these models,
our results should only be
viewed as a first pass through this class of models.  It is possible (but by no means certain) 
that more generic vacua
in this class (based on compactification 
manifolds which have more complicated geometry, or based on studies of other moduli) 
would
yield different results.  More generally, flux potentials in other classes of vacua
may well have broader distributions of the relevant physical quantities
for inflation, allowing one to tune fluxes to achieve inflation.\footnote{Very concrete
reasons to expect that
the distributions are broader in IIB flux vacua are described in \S6.2\ of \cite{DeWolfe}, for
instance.}

The rest of this paper is organized as follows. 
Section~\ref{StringTheory} is a basic review of string compactification aimed at the non-specialist.
Here we review the process of moving from the 10-dimensional theory
to the 4-dimensional theory in fairly simple terms, showing
how the 4-dimensional picture has the ingredients of inflation 
(gravity as well as kinetic and potential energies for scalar fields).
We then show how the familiar slow-roll conditions for inflation become slightly generalized due to the non-standard kinetic terms from string theory.
In Section~\ref{Models}, we present and analyze the three explicit models analytically and numerically.
We summarize our conclusions in Section~\ref{Discussion}.

\section{String Theory and Dimensional Reduction}\label{StringTheory}
In this Section we give a gentle introduction and review of the study of compactification 
in string theory, with a focus
on the ingredients that are relevant for the specific models we will investigate in Section \ref{Models}. 
Much more complete and technical reviews are given in \cite{evareview,Grana,Douglas,Lust_1,Lust_2,Lust_3}, 
while a qualitative introductory review appears in \cite{Denef}.
We begin by mentioning the basic ingredients of type IIA string theory, with a focus on fluxes.

\subsection{Supergravity}\label{Supergravity}

String theory is believed to be a consistent theory of quantum gravity.
One curious feature of the theory is that this {\em consistency} suggests a special role for 10 dimensions,
where consistent string theories are in correspondence with the so-called
``maximal supergravities".\footnote{Here we are referring to the $\mathcal{N}=1$ and $\mathcal{N}=2$  theories in 10-dimensions, where $\mathcal{N}$ is the number of supersymmetries.
}
Furthermore, a remarkable feature of the theory is that its dynamics in 10 dimensions
can be {\em derived}, rather than {\em guessed}, by demanding consistency. 
For comparison, consider the familiar
case of a charged point particle moving in a background curved space-time $g_{\mu\nu}$
with a background electromagnetic field strength $F_{\mu\nu}$. 
There is no reasonable way to uniquely derive the dynamical equations governing the time-evolution of
$g_{\mu\nu}$ and $F_{\mu\nu}$ from any consistency arguments about the behavior of
the point particle. However, in the case of the string, this is precisely what happens.

In this paper, we will focus on what is known as the type IIA string theory.
It can be derived that part of the 10-dimensional action governing gravity and the field strength of the string in this theory is \cite{Polchinski}
\begin{eqnarray}
S \amp=\amp\frac{1}{2 \kappa_{10}^2}\int d^{10}\!x\sqrt{-g_{10}}\,
e^{-2\phi}\nonumber\\
\amp\amp\,\,\,\,\,\,\,\,\,\,\,\,\,\,\,\,\,\times\Big{(}R_{10}+4(\partial_{\mu}\phi)^2
-\frac{1}{2} F_{\mu\nu\rho}F^{\mu\nu\rho}\Big{)}.\label{gravity1}
\end{eqnarray}
Here $R_{10}$ is the 10-dimensional Ricci scalar, $\phi$ is a scalar field known as the ``dilaton'', 
and $F_{\mu\nu\rho}$ is a generalized
electromagnetic field strength; it carries 3 indices (making it a so-called 3-form) 
since it is sourced by the 
$(1+1)$-dimensional string, just like the familiar electromagnetic field strength $F_{\mu\nu}$ 
carries 2 indices (a 2-form)
since it is sourced by a $(0+1)$-dimensional point particle.
The overall pre-factor sets the gravitational strength in 10 dimensions $\kappa_{10}^2=8\pi G_{10}$.
It is related to the so-called Regge-parameter $\alpha'$ (units of length-squared) by $2\kappa_{10}^2=(2\pi)^7\alpha'^4$.
The inverse of the Regge-parameter is the string tension ($= 1/2\pi\alpha'$);
the tension of a string is an absolute constant and 
is analogous to the mass of a particle.
Furthermore, $\alpha'$ is related to the string length by (in our convention) $l_s=2\pi\sqrt{\alpha'}$. 
Later we will see that it is convenient to measure a number of dimensionful
quantities in units of $l_s$.
Table \ref{deftable} provides a hopefully useful dictionary of key string theory notation and the symbols used in this paper, including a summary of the above.

Let us summarize: In eq.~(\ref{gravity1}) we see that the 10-dimensional universe of string theory
contains gravity and a field strength $F_{\mu\nu\rho}$, and that they appear in the same 
way as gravity and electromagnetism do in 4 dimensions. 
We further note that there exists the dilaton $\phi$ which is non-minimally coupled to gravity.
Because the coefficient of $F_{\mu\nu\rho}F^{\mu\nu\rho}$ is proportional to $e^{-2\phi}$,
one identifies $g_s\equiv e^{\phi}$ as the string coupling; it is the string loop expansion parameter analogous to $e$ in electromagnetism.

Now we must mention some other features of string theory in 10 dimensions
that we did not include in eq.~(\ref{gravity1}).
First of all, it turns out that the 3-form $F_{\mu\nu\rho}$ (which we will later denote simply $F_3$, and which is often denoted $H_3$ in the string literature)
is {\em not} the only field strength that appears in string theory. 
Rather, there are also various other
fields; various so-called $p$-forms with $p$ indices, where $p$ takes various integer values, 
and whose interactions are also uniquely determined by consistency.  In addition, there are extended objects of various dimensionality in the theory known as Dirichlet branes and orientifold planes, which are charged under these $p$-forms \cite{Dbranes}.
Also, 
there are fermions which
give rise to a collection of terms to be added to eq.~(\ref{gravity1}), since we are describing a supersymmetric theory, but we have set their values to zero here.
We focus on cosmologies that have maximal space-time symmetry (Minkowski,  anti de Sitter, de Sitter) which means that the vacuum expectation values of the fermion fields must vanish.
Finally, what appears in eq.~(\ref{gravity1}) is only the first
term in a perturbative expansion in powers of $\alpha'$ and $g_s$. For length scales large compared
to the string length $l_s$ and for small $g_s$ we can ignore such corrections; this is known as the supergravity approximation.

\subsection{Compactification and Fluxes}\label{Compactification}

\subsubsection{Calabi-Yau manifolds}

Of most interest to us is what this theory predicts in 4 dimensions.
Currently there is no background independent formulation of string theory, so the {\em compactification} of the 10-dimensional geometry
to 4 large dimensions is specified by hand
and is not unique.
The most commonly studied compact spaces are Calabi-Yau manifolds 
(e.g., see \cite{Douglas,Polchinski} for a technical definition).
They are useful for at least two reasons: Firstly, Calabi-Yau manifolds preserve some 
unbroken supersymmetry which allows for better computational control. 
Secondly, and most importantly for us,
Calabi-Yau manifolds are spaces that possess a metric 
that is Ricci flat (has $R_{\mu\nu}=0$ like any vacuum metric). This is very convenient in finding solutions to the 10-dimensional equations of motion.
The simplest example is that of the torus $T^6$, which is not only Ricci flat, but also flat (with vanishing Riemann tensor). 
In this paper we will focus on this space ($T^6$),
since this has been studied the most intensely in the literature. 
It is also a very useful pedagogical device, and we will make some comments below on the connection
of our results to more general compactifications.

\subsubsection{Orbifolds and orientifolds}

Although understanding them in detail is not central to following our examples below, let us briefly
mention orbifolding \cite{orbifold} and orientifolding \cite{orient_1,orient_2,orient_3}, 
two technical operations that string theorists perform
on the compact space, since they occur in all the models we will investigate.
It often proves to be important to reduce the number of points in a manifold by declaring some of them identical.
In technical jargon, one forms the quotient space with some finite symmetry group of the manifold, 
for example $T^6/\mathbb{Z}_p$, where $\mathbb{Z}_p$ is the group of integers modulo $p$.
The specific $\mathbb{Z}_p$ symmetry is model dependent.
This defines a so-called orbifold. Certain toroidal orbifolds are of interest since they are a special singular limit
of some (non-toroidal) Calabi-Yau.\footnote{E.g., the orbifold $T^4/\mathbb{Z}_2$ is a limit of the Calabi-Yau K3.} 
Also, by performing additional discrete operations one can form what is known as an orientifold.
It is related to forming unoriented strings out of oriented strings. It will be important for us in what follows that at fixed points of the group action on the internal manifold, in orientifold models, one gets so-called {\it orientifold~planes} (or O-planes). These O-planes provide a negative contribution to the vacuum energy (see ahead to eq.~(\ref{Oplane})) and this is important for stabilization.

Both of these operations, orbifolding and orientifolding, serve the purpose
of allowing for chiral fermions and reducing the amount of supersymmetry in 4 dimensions.
For suitable choices of symmetry groups, there 
can however be a residual amount of supersymmetry in 4 dimensions.\footnote{
For example, in commonly studied orientifolds of Calabi-Yau manifolds,
the $\mathcal{N}=2$ theory 
in 10 dimensions becomes an $\mathcal{N}=1$ theory in 4 dimensions in type II string theory.}

 
\subsubsection{Moduli}
 
In general, there
are scalar fields characterizing the size and shape of any compact manifold: K\"ahler moduli
(roughly specifying size) and complex structure moduli (roughly specifying shape).
Table \ref{deftable} summarizes all fields whose dynamics we will keep track of in 4 dimensions:
besides gravity $g_{\mu\nu}$, we have ``geometric" moduli: a dilaton $\phi$, 
K\"ahler moduli (also known as ``radions''), and complex structure moduli. 
In addition, each of these geometric moduli
are accompanied by a field that is generically referred to as an ``axion''. 
The reason these are called axions is not important here, but suffice it 
to say that they are all pseudo-scalars and
some are coupled to a generalized ${\bf E}\cdot{\bf B}$ term in the
action, reminiscent of the axion of quantum chromodynamics (see, e.g., \cite{Conlon,Svrcek}).
If we denote the various geometric moduli by $\gmog_i$ 
($i=1,2,...$) and the axions by $\gmoa_i$, then these two degrees
of freedom can be put into a complex pair $\gmo_i = \gmoa_i +i\, \gmog_i$.\footnote{In the string literature, there are many symbols used for the different moduli, such as $T$, $U$, $v$, etc, but we will just use the common notation $\gmo_i = \gmoa_i +i\, \gmog_i$ for all moduli.}
We will see that this construction of forming a complex scalar is quite useful.\footnote{In fact this construction is integral to ${\cal N}=1$
supersymmetric models, where such pairs are unified in a chiral multiplet, a representation of the
supersymmetry algebra.}
We will group all these complex fields into a single vector $\boldsymbol{\psi}$, which will act as our complex inflaton vector. When separated into its real components, we denote this 
$\boldsymbol{\phi}$; our real inflaton vector.

\subsubsection{Fluxes and potential energies} 

Strictly speaking, moduli are defined as those scalars that have vanishing potential. 
Without including any extra ingredients (such as field strengths), the above mentioned fields would indeed be massless
and free. This is very problematic. For example, if the radions are freely propagating fields
then the size of the compact space could take on any value, including
unacceptably large values. 
Indeed, there are constraints from 5-th force experiments showing that
these fields must be stabilized with a large effective mass (moderately large compared
to the inverse millimeter scale to which gravity has been tested, and huge compared to today's Hubble scale), \ie, that there must be contributions to the potential energy density of the form $m_i^2 \gmog_i^2$ (where
the coefficients $m_i$ are large).
Furthermore, we are interested in whether any of these scalars could be the inflaton.
Since a free field by definition is one that does not feel a potential, it cannot possibly drive inflation.

However, an important feature of string theory is the 
existence of various field strengths,
and these induce interactions for the moduli. We have already introduced the field strength 
$F_{\mu\nu\rho}$, hereafter abbreviated $F_3$.
We will focus on what is known as type IIA string theory in this paper, in which there are
also forms with even numbers of indices, such as $F_2$ and $F_4$, but more general forms $F_p$ occur in other models.
In our 3+1 large dimensions, Lorentz invariance prevents any cosmological
field strengths, however such restrictions do not apply to the components of $F_p$ in the
compact space.\footnote{In the presence of field strengths in the compact space,
the Ricci tensor on the compact space is in general non-zero, and so the space is strictly no longer Calabi-Yau. However, when the moduli masses are light compared to the inverse size of the compact space (the ``Kaluza Klein scale") then this {\em back-reaction} is small, and we can continue to treat the compact space as Calabi-Yau. This is a property of the models we study.
}
 Assuming $p\leq 6$, then the fields satisfy
\begin{equation}
\frac{1}{l_s^{p-1}}\int F_p = f_p,
\label{flux}\end{equation}
where the integral is over some $p$-dimensional internal manifold of the compact space.
Such integrals appear when we compactify the theory. 
Here $f_p$ is an integer, corresponding to a generalized Dirac charge quantization condition. 
These quantized integrals of the field strengths are known as `fluxes'.
They correspond to wrapped field lines in the compact space.
Such fields can be thought of as being sourced by generalized electric and 
magnetic charges provided by the various branes of the theory \cite{Douglas}. 
Note, however, that this is just an incomplete analogy, 
since the fluxes we are referring to here
thread topologically non-trivial internal submanifolds of the compact space; therefore, Gauss'
law does not require charges to source the flux.
(There are, however, other space-filling branes in the theory, which will appear in the models).
Since there is an energy cost associated with deformations of the compact space in
the presence of field strengths, these fluxes induce a potential energy $V=V(\boldsymbol{\psi})$ for the moduli.
This potential is necessary for stabilization, and we will investigate if it can
also drive inflation. 

To get an idea of the form that these energies will take, consider a generalized electric field ${\bf E}_p$ set up by a stationary source; a point source for ${\bf E}_2$, a string for ${\bf E}_3$, a membrane
for ${\bf E}_4$ etc. If there are $f_p$ units of charge contained in a compact space of size $r$, then
the energy density is given roughly by (ignoring factors of $l_s$)
\begin{equation}
|{\bf E}_p|^2 \sim \frac{f_p^2}{r^{2 p}},
\label{e2}\end{equation}
which reduces to the familiar result $f_2^2/r^4$ for a point source.
The total energy density will involve a sum of such terms. eq.~(\ref{energy}) below illustrates the form this takes
when expressed in a 4-dimensional action.

\subsection{The 4-Dimensional Action and Slow-Roll}\label{fourdaction}

Here our aim is to move from the 10-dimensional theory to the 4-dimensional theory.

\subsubsection{Integrating out the compact space}

To understand the 4-dimensional action, let us begin by focusing on the gravity sector.
For simplicity, we will assume that the 10-dimensional metric is in block diagonal form: $g_{AB}^{10}=\mbox{diag}(g_{\mu\nu}^4,g_{ab}^6)$,
i.e., that it separates into a piece governing the 4 large space-time dimensions and a piece 
governing the 6 compact dimensions.
Furthermore, let us assume that 
the Ricci scalar is
independent of the compact co-ordinates (the usual
assumption in Kaluza-Klein models), so that we can integrate over $d^6x$. This assumption is valid when one
considers an (approximately) ``unwarped compactification,'' as we do in this paper.
The relevant piece in the action is the first term in eq.~(\ref{gravity1}),
involving the 10-dimensional Ricci scalar $R_{10}$ and the dilaton (scalar field). This gives
\begin{eqnarray}
 \int d^{10}x\sqrt{-g_{10}}\,e^{-2\phi}R_{10} 
 = 
 \int d^{4}x\sqrt{-g_4}\,\mbox{Vol}\,e^{-2\phi}R_{10}
\label{noncan}\end{eqnarray}
where Vol is the 6-dimensional volume of the compact space. The particular form of Vol is model dependent,
and the relationship between the 10-dimensional Ricci scalar $R_{10}$ and the
4-dimensional Ricci scalar $R_4$ is also model dependent through the form of the metric on the
compact space. However, for this class of models it is true that
\begin{equation}
R_{10}=R_4 + \mbox{func}(\mbox{compact space fields})
\end{equation}
Since both the volume Vol and the dilaton $\phi$ are allowed to be dynamical, the action in eq.~(\ref{noncan}) is evidently not in canonical form. In order to bring the  action into canonical form,
we introduce the so-called Einstein metric $g_{\mu\nu}^{E}$, which is defined via a conformal transformation as
\begin{equation}
g_{\mu\nu}^{E}\equiv \frac{\mbox{Vol}\,e^{-2\phi}}{\mpl^2\,\kappa_{10}^2} g_{\mu\nu}^4
\label{Einconf}\end{equation}
where $g_{\mu\nu}^4$ is the 4-dimensional string metric -- the metric in the ``string frame" of eq.~(\ref{gravity1}).
In this transformation we have introduced the (reduced) Planck mass $\mpl=1/\sqrt{8\pi G}\approx2\times 10^{18}$\,GeV,
where $G$ is the 4-dimensional Newton constant.
The gravity sector, written in terms of the corresponding Einstein Ricci scalar $R_E$, then appears in canonical form
\begin{eqnarray}
&&\frac{1}{2\kappa_{10}^2}\int d^{10}x\sqrt{-g_{10}} e^{-2\phi} R_{10} \nonumber \\
&&\,\,\,\,\,\,\,\,\,\,\,\,\,\,\,\,\,\,
= \int d^4 x\sqrt{-g_E} \left(\frac{1}{16\pi G}R_E + \cdots \right)
\end{eqnarray}
and this is referred to as the ``Einstein frame''.

Let us clarify a feature of the conformal transformation. Recall that during inflation, both Vol and $\phi$ may be dynamical, since they in fact depend on the inflaton vector $\psi$.
Well after inflation, we expect these fields to be stabilized at some fixed values: 
$\langle \mbox{Vol} \rangle$ and $\langle\phi\rangle$. At such values we require that the conformal transformation in eq.~(\ref{Einconf}) be the identity transformation. 
This implies a relationship between $\kappa_{10}$, $\langle\mbox{Vol}\rangle$, $\langle\phi\rangle$, and $\mpl$:
\begin{equation}
\mpl^2 = \frac{\langle\mbox{Vol}\rangle\,e^{-2\langle\phi\rangle}}{\kappa_{10}^2}. 
\end{equation}
Note that the Planck mass is defined in terms of the fields at their {\em stabilized values}, so it is a constant.
Since it is natural for $\langle\mbox{Vol}\rangle$ to be given in units of $l_s^6$ and since $\kappa_{10}^2=l_s^8/4\pi$,
one can in fact use this equation to determine the string length $l_s$ in terms of the Planck length for any particular model.

\subsubsection{Kinetic energy}

In general, the kinetic energy of the moduli is not in canonical form. Recall that the
moduli are a combination of not only the dilaton, but also the K\"ahler and complex structure moduli
which describe the size and shape of the particular Calabi-Yau compact space. An important property
of every Calabi-Yau (related to the underlying supersymmetry it preserves)
is that there exists a scalar function $K$ of the moduli, known as the K\"ahler potential,
whose Hessian matrix is the metric on moduli space:
\begin{equation}
K_{i\bar{j}} \equiv K,_{i\bar{j}} = \frac{\partial^2 K}{\partial\psi^i\partial\psi^{\bar j}}, 
\end{equation}
where $\psi^i$ is a generic name for a complex modulus (axion partnered with geometric moduli),
and barred variables denote the complex conjugate. Contracting space-time derivates of
the moduli with this metric gives the kinetic energy in the Einstein frame:
\begin{equation}
T = - \mpl^2 K_{i\bar{j}}\partial_\mu\psi^i\partial^{\mu}\psi^{\bar{j}}.
\end{equation}
There is an implicit sum over all $i,j$. Since we shall deal with dimensionless K\"ahler potentials, the factor of $\mpl^2$ is necessary on dimensional grounds.
In the limit in which we shall work,
$K$ does {\it not} depend on fluxes, which means that there is one K\"ahler potential
for each of the three models that we will investigate, applicable to any vacuum in their respective landscapes. 
In other words, fluxes affect only the potential energy, not the kinetic energy.

\subsubsection{Potential energy}

In addition to the K\"ahler potential $K$, there exists another object which contains information
about the particular compactification.
For supersymmetric compactifications, of the type we will focus on, this object is the so-called superpotential $W$. 
$W$ is a complex analytic function of the complex moduli $\psi^i$ and
also depends on the fluxes. If
we turn on fluxes, then there are in general energies induced associated with distortions of the
compact space and displacements of the dilaton and axions. 
These interactions are all contained in $W$. 
The potential energy term in the 4 dimensional Lagrangian density is given by
\begin{equation}
V=e^K\!\left(D_i W K^{i\bar{j}}\overline{D_j W} - 3 |W|^2\right),
\label{SUGRA}\end{equation}
which is sometimes referred to as the supergravity formula.
Here $D_i W \equiv \partial_i W+W\partial_i K$ and $K^{i\bar{j}}$ is the matrix inverse of $K_{i\bar{j}}$.
Again there is an implicit sum over all $i,j$.
A reader familiar with supersymmetry in 4 dimensions will recognize that these are the 
so-called ``F terms" and that there are no so-called ``D terms". 

In order to develop some intuition, we would like to get an idea of the typical form of $V$.
In eq.~(\ref{e2}) we estimated the energy density established by a $p$-form field strength, but must now take into account the integration over the compact space and the conformal transformation.
For a compact space of size $r$, the contribution to $V$ is given roughly by (ignoring factors of $l_s$)
\begin{eqnarray}
&& \Delta V \sim f_3^2\frac{e^{2\phi}}{r^{12}}\,\,\,\,\,\,\,\,\,\,\,\mbox{for $F_3$,}\nonumber\\
&& \Delta V \sim f_p^2\frac{e^{4\phi}}{r^{6+2 p}}\,\,\,\,\,\mbox{for $F_{p\ne 3}$.}
\label{energy}\end{eqnarray}
We also mention an estimate for the contribution to $V$ from $N_1$ D6-branes and $N_2$ O6-planes
\begin{eqnarray}
&& \Delta V \sim N_1 \frac{e^{3\phi}}{r^9}\,\,\,\,\,\,\,\,\,\,\mbox{for D6-branes,}\nonumber\\
&& \Delta V \sim -N_2 \frac{e^{3\phi}}{r^9}\,\,\,\,\,\mbox{for O6-planes.}
\label{Oplane}\end{eqnarray}
Note that the O6-plane makes a negative contribution.

In the models studied, when calculating the potential from the supergravity formula $V$, we will for simplicity work in units where
the string length $l_s=1$.
This must be rescaled in order to obtain the potential in Einstein frame $\VE$ in conventional units:
\begin{equation}
\VE  = \frac{\mpl^4}{4\pi} V.
\label{VEinstein}\end{equation}
(This comes from: $\mpl^4\kappa_{10}^2=\mpl^4\,l_s^8/4\pi=\mpl^4/4\pi$.)
In Sections \ref{DGKT} -- \ref{bu} we will just refer to $V$, rather than $\VE$.

\subsubsection{Putting it all together}

Altogether, the effective 4-dimensional action in the Einstein frame is a familiar sum of a gravity term, kinetic energy, and potential energy, i.e.,
\begin{eqnarray}
S \amp=\amp \int d^4x\sqrt{-g_E}\nonumber\\
\amp\amp
\times\left[\frac{1}{16\pi G}R_E - \mpl^2K_{i\bar{j}}\partial_\mu\psi^i\partial^{\mu}\psi^{\bar{j}} - \VE(\boldsymbol{\psi})\right]\!.\,\,\,\,\,\,\,\,\,\,\,
\label{actionK}\end{eqnarray}
Keeping the kinetic term general, the Euler-Lagrange equations of motion 
for a flat Friedmann-Robertson-Walker (FRW) 
universe $g_{\mu\nu}^E=\mbox{diag}(-1,a(t)^2,a(t)^2,a(t)^2)$
are in terms of $K$ and its derivatives:
\begin{eqnarray}
&& \ddot{\psi}^i+3H\dot{\psi}^i+\Gamma^i_{jk}\dot{\psi}^j\dot{\psi}^k+K^{i\bar{j}}\VE_{,\bar{j}}/\mpl^2 = 0,\label{fieldeqn} \\
&& H^2 = \frac{8\pi G}{3}\left[\mpl^2K_{i\bar{j}}\dot{\psi}^i\dot{\psi}^{\bar{j}} + \VE(\boldsymbol{\psi})\right], \label{Hubbleeqn}
\end{eqnarray}
where $H=\dot{a}/a$ is the Hubble parameter and $\Gamma^i_{jk}=K^{i\bar{n}}K_{jk,\bar{n}}$ is the Christoffel symbol on moduli space.

To investigate inflation, we must compute the slow-roll parameters. 
The first slow-roll condition is that the kinetic energy in (\ref{Hubbleeqn}) should be small compared to the potential energy when 
the acceleration term in (\ref{fieldeqn}) is negligible.
The second condition is that the acceleration is and remains small along the slow-roll direction (quantified by differentiating 
and demanding self-consistency).
These two conditions define the two slow-roll parameters $\epsilon$ and $\eta$ and correspond to the requirements that
$\epsilon<1$ and $|\eta|<1$.
We find that
\begin{eqnarray}
&& \epsilon=\frac{K^{i\bar{j}}V_{,i}V_{,\bar{j}}}{ V^2} \left(= \frac{g^{ab}V_{,a}V_{,b}}{2V^2}\right),\label{SReps}
\\ && \eta = \mbox{min eigenvalue}\left\{\frac{g^{ab}\left(V_{,bc}-\Gamma^d_{bc}V_{,d}\right)}{V}\right\}.
\label{SReta}\end{eqnarray}
Here $\eta$ (and $\epsilon$) is written in terms of the metric $g_{ab}$ 
governing real scalar fields $\phi^a$:
$K_{i\bar{j}}\partial_\mu\psi^i\partial^\mu\psi^{\bar{j}}=\frac{1}{2}g_{ab}\partial_\mu\phi^a\partial^\mu\phi^b$, with $\phi^{2i-1}=a^i$ and $\phi^{2i}=b^i$.
Note that we can use either $V$ or $\VE$ in the slow roll conditions, as $V\propto\VE$ (see eq.~(\ref{VEinstein})).
In regions where inflation occurs, these three functions ($V,\,\epsilon,\,\eta$) can be used to predict several cosmological parameters, 
as detailed in Appendix D. A comparison between the theoretical predictions and observational data provides a precision test of the model.

\subsection{de Sitter Vacua}\label{deSitter}

There are two good reasons to want de Sitter ``vacua", namely that
there are at least two eras of our universe that are approximately de Sitter;
during inflation, which exhibits slow-roll, and at late times,  which appears consistent with a positive cosmological constant.
This article is focussed on the former epoch.
If we have a region in moduli space that is de Sitter, namely a region
in which the gradient of the 4-dimensional potential $V$ (from eq.~(\ref{actionK})) is zero with $\VE>0$, we are a significant step closer to realizing inflation. In such a region (which may be a single point) we at least know that
the first slow-roll parameter $\epsilon=0$, although we may still face the so-called $\eta$-problem 
\cite{etaprob} (for a recent discussion in the string theory context, see \cite{etarecent}).

Let us make some comments about supersymmetric vacua and anti de Sitter (AdS) space.
The condition for supersymmetry (SUSY) is that the covariant derivative of the superpotential vanishes, i.e.,
\begin{equation} 
D_i W = \partial_i W+W\partial_i K = 0
\label{SUSY}\end{equation}
for all $i$. It is simple to show from eq.~(\ref{SUGRA}) that at any such
SUSY point the supergravity potential $V$ is stationary. This constitutes some ``vacuum" (stable or unstable) of the theory.
At such a point the potential is
\begin{equation}
V_{\mbox{\tiny{SUSY}}} = -3 e^K|W|^2,
\end{equation}
so we see it is necessarily non-positive. It may happen that $W=0$, corresponding then to Minkowski space.
But what is \emph{much more common} is for $W\ne 0$, corresponding then to anti-de Sitter space.

In fact it  is known that any vacuum 
that is supersymmetric (in supergravity, superstring theory, M-theory, etc) is necessarily non de Sitter.
But it may be the case that in such a model, a non-SUSY minimum in the space of scalar field expectation values
is de Sitter (spontaneously broken SUSY).  Achieving this is not easy, as described by various `no-go theorems'. 
In particular, under  
mild assumptions on the 
nature of 
the compact space (namely that it is non-singular etc.),   
one can show that inclusion of fluxes alone does not allow one to find any de Sitter vacua \cite{Gibbons,Nunez}.

There are, however, other structures besides fluxes in string theory, e..g, D-branes and O-planes.
In \cite{Giddings} it is shown that the argument of 
\cite{Nunez} may be extended to include most forms of D-branes, but it cannot be 
extended to include O-planes. 
The energies associated with such structures were described in eqn. (\ref{Oplane}).  Recent work has shown that 
the realization of de Sitter vacua is possible with such ingredients, as in the constructions of 
\cite{Eva}.  However, we do not find any de Sitter vacua in the simple IIA models we study.

Given that we are considering models with AdS vacua,
what is the implication for inflation? Suppose that the inflaton eventually settles down to some such AdS vacuum.
There the potential has a (negative) value that we will call the `cosmological constant'.
This may well compromise any chance to obtain many e-folds of inflation, which
requires $\VE>0$. However, we can imagine a priori a scenario where this is \emph{not} catastrophic
for inflation: We will see later that we have fluxes that can be used to dial the cosmological constant 
toward zero. Hence the depth of AdS space can be tuned very small. Furthermore,
well away from the SUSY vacuum there are regions in moduli space were the potential is large and \emph{positive}.
Then, as long as $\VE$ during inflation is much greater than the depth of the AdS minimum, it is plausible
that inflation could be realized. 

At the end of inflation one should in principle enter the radiation era.
Normally this occurs through the decay of the inflaton to various fields including the standard model particles. However, the standard model is not contained in the models that we investigate,
so this is an issue that we do not tackle. Furthermore, we do not address the late-time problem of the smallness of the (positive) cosmological constant.
(A popular explanation of the smallness of the cosmological constant appeals to the existence 
of exponentially many vacua realizing different vacuum energies,
see \cite{Bousso}).

\section{Type IIA Models}\label{Models}

Here we investigate the cosmology of three explicit models. 
Many choices of compactification are possible. However, the
torus is flat and is perhaps the most well studied compact manifold
in the literature, so we will focus on (orbifolds and orientifolds of) this. 
We will investigate the resulting inflation picture for three explicit models: 
DGKT \cite{DeWolfe}, VZ \cite{Villadoro}, and IW \cite{Ihl}. 
In this Section 
we use units $l_s=1$.

\subsection{Diagonal Torus Models}\label{DiagonalTorus}

For clarity, let us describe the properties of the K\"ahler potential and the
slow-roll conditions in more detail for a particular class of examples.
The first two torus models to be discussed have the property that the 
K\"ahler potential is the logarithm of a product of moduli. Writing 
\begin{equation}
\gmoi=\gmoai+i\gmogi
\end{equation} 
for all moduli\footnote{More precisely, we study all moduli that arise from metric deformations of the torus, the dilaton, and their superpartners.  We neglect
so-called ``twisted moduli" or ``blow-up modes'' 
originating from singularities of the orbifold group action, though we briefly discuss them in 
Section \ref{bu}.},
we have
\begin{equation}
K = -\ln\left(\prod_i\gmogi^{n_i}\right)+\mbox{const}
\label{Kprod}\end{equation}
where $n_i$ are $\mathcal{O}(1)$ integers (e.g., in the VZ model described below, there are 7 moduli 
($i=1,\ldots,7$) with $n_i=1$). The kinetic energy for such models is then
\begin{equation}
T = - \mpl^2\sum_i \frac{n_i}{4} \frac{(\partial_\mu \gmoai)^2+(\partial_\mu \gmogi)^2}{\gmogi^2}
\label{actioncan}\end{equation}
In this case, the equations of motion for the moduli are
\begin{eqnarray}
0&=&\ddot{\gmog}_i+3H\dot{\gmog}_i+\frac{\dot{\gmoa}_i^2 - \dot{\gmog}_i^2}{\gmogi}
+\frac{1}{\mpl^2}\frac{2\gmogi^2}{n_i}\frac{\partial \VE}{\partial \gmogi},\,\,\,\, \\
0&=&\ddot{\gmoa}_i+3H\dot{\gmoa}_i -2\frac{\dot{\gmoa}_i\,\dot{\gmog}_i}{\gmogi} 
+\frac{1}{\mpl^2}\frac{2\gmogi^2}{n_i}\frac{\partial \VE}{\partial \gmoai}.
\end{eqnarray}
The first slow-roll parameter for inflation then takes the form
\begin{eqnarray}
\epsilon = \frac{1}{V^2}\sum_i\frac{\gmogi^2}{n_i}\left[\left(\frac{\partial V}{\partial \gmoai}\right)^2+
\left(\frac{\partial V}{\partial \gmogi}\right)^2\right].
\label{epsilon}\end{eqnarray}

It is important to take note of the form of the K\"ahler potential; it is independent
of all axions (general fact). Hence if we shift an axion by a constant, it has no effect on
the kinetic energy. Also, if we rescale $\gmoi$ by a real number, the kinetic energy is
also unchanged. In short,
\begin{equation}
\gmoai \to d_i\gmoai + c_i,\,\,\, \gmogi \to d_i\gmogi
\end{equation}
leaves the kinetic energy unchanged for any constants $c_i, d_i\in\mathbb{R}$.
In turn, the form of the slow-roll parameters $\epsilon$ and $\eta$ are unaffected.
These shift and scaling symmetries
allow one to eliminate some flux parameters that appear in the superpotential.
In the first model, we will see that these symmetries allow 
all fluxes to be absorbed into an overall multiplicative factor,
while in the second and third models we will have one additional non-trivial flux parameter to dial.
There will in general be ambiguities associated with positive/negative values
of the fluxes that one should keep careful track of.

In order to emphasize that the Lagrangian in 4 dimensions is reminiscent of that in standard inflation, 
let us perform a field redefinition for the simple case where we ignore the axions 
and focus on $\VE(\gmogi)$. By defining
\begin{equation}
\phi_i \equiv \sqrt{\frac{n_i}{2}}\,\mpl\log\gmogi,
\end{equation}
the kinetic energy is put in canonical form and
the action in eq.~(\ref{actionK}) becomes
\begin{eqnarray}
S \amp=\amp \int d^4x\sqrt{-g_E}\nonumber\\
\amp\amp
\times\left[\frac{1}{16\pi G}R_E - \sum_{i}\frac{1}{2}(\partial_{\mu}\phi_i)^2 - \VE\Big{(}e^{\frac{\sqrt{2}\,\phi_i}{\sqrt{n_i}\mpl}}\Big{)}\right]\!.\,\,\,\,\,\,\,\,\,\,\,
\label{actionSimple}\end{eqnarray}
Note that the argument of $\VE$ is now an exponential.
The first slow-roll parameter then takes the canonical form for multi-field inflation:
\begin{equation}
\epsilon = \frac{\mpl^2}{2}\frac{|\nabla_{\!\phi} V|^2}{V^2}.
\end{equation}
We point out that this is only true when ignoring the axions and relies upon the assumed simple form of the K\"ahler potential. 

\subsection{The Model of DeWolfe, Giryavets, Kachru, and Taylor (DGKT)}\label{DGKT}
In May 2005, DeWolfe et.al \cite{DeWolfe} (DGKT) found an explicit {\em infinite} class of stable vacua
in type IIA string theory. In their model they found that all moduli are stabilized
by including the 3-form, 4-form (and less importantly the 2-form) fluxes of type IIA, and also
including a 0-form flux. The 0-form flux plays the role of a mass-term in the theory.
Its presence induces several extra pieces into the action, which can only be derived from so-called M-theory.
This framework is known as ``massive type IIA supergravity".

Starting from the torus, they built the orbifold $T^6/\mathbb{Z}_3$ 
and projected it to the orientifold $T^6/\mathbb{Z}_3^2$. 
In addition, they introduced a static $(6+1)$-dimensional plane that carries charge (an O6-plane),
for the purpose of satisfying a constraint known as a tadpole condition.
We will focus here on reporting the salient features
of the geometry after these technical operations have been performed; the reader is referred to 
the original paper \cite{DeWolfe} for details.

The torus of DGKT takes on essentially the simplest possible form (we will see more complexity in the
later models of VZ and IW). The orbifolding and orientifolding act to 
reduce the number of degrees of freedom of the metric on the compact space
to just 3. Here we neglect the moduli which arise at the orbifold fixed points (the so-called
``blow up modes" or ``twisted sector" moduli).\footnote{We will briefly discuss the proper
inclusion of the blow-up modes in \ref{bu}.  They alter the discussion in various important
ways, but do not at first sight seem to change our conclusions.}
These 3 are the (untwisted) 
K\"ahler moduli of the theory.  
There are no complex structure structure moduli left.
The metric on the compact space and the volume are given by: 
\begin{eqnarray}
(ds^2)_6 \amp =\amp \sum_{i=1}^3\gamma_i \left[(dx^i)^2+(dy^i)^2\right], \\
\mbox{Vol} \amp = \amp \int_{T^6/\mathbb{Z}_3^2}
d^6x\sqrt{g_6}=\frac{1}{8\sqrt{3}}\gamma_1\gamma_2\gamma_3=\mokg_1\mokg_2\mokg_3.\label{VolDGKT}\,\,\,\,\,\,\,\,\,\,\,\end{eqnarray}
Here the elements of the metric are called $\gamma_i$. The volume is proportional to the determinant 
of the square root of the metric ($\gamma_1\gamma_2\gamma_3$), 
the factor of $1/8\sqrt{3}$ comes from performing the peculiar integration over $T^6/\mathbb{Z}_3^2$,
but is not important for us. What is important is the identification of the good K\"ahler co-ordinates $b_1,\,b_2,\,b_3$ 
whose product is the volume\footnote{In the DGKT paper: $\mbox{Vol}=\kappa\,\mokg_1\mokg_2\mokg_3$, with $\kappa=81$. By rescaling $\mokg_i\to\kappa^{-1/3}\mokg_i,\,i=1,2,3$ we obtain eq.~(\ref{VolDGKT}) and $\kappa$ is eliminated.}  (up to a prefactor, they are just the components of the metric).

Let us summarize the moduli of this model.
As mentioned, all complex structure moduli are projected out by the orientifolding,
leaving 4 moduli: 3 K\"ahler moduli $\moki = \mokai+i \mokgi,\,i=1,2,3$ 
and an axio-dilaton $\moa=\moaa+i \moag\,\,(\moag=e^{-\phi}\sqrt{\mbox{Vol}}/\sqrt{2})$.\footnote{The
canonical model-independent axion is $\xi=2a_4$.} We note that the axio-dilaton appears in
the compactification, not through an explicit appearance in the compact metric, but through its direct appearance
in the action, as discussed in Section \ref{StringTheory}.

The K\"ahler potential takes on the form promised in Section \ref{DiagonalTorus}, namely
the logarithm of the product of geometric moduli. All that is left is to specify the values of $n_i$
and the constant. One finds that
\begin{equation}
K = -\ln\left(32\,\mokgo\,\mokgt\,\mokgth\,\moag^4\right).
\label{DGKTKahler}\end{equation}

The superpotential is set by the interactions:
DGKT turned on fluxes coming from $F_3,\,F_2,\,F_4$, and a zero form $F_0$, a so-called mass term,
as well as an $F_6$. By studying the work of Grimm and Louis \cite{Grimm} they find
\begin{eqnarray}
W \amp =\amp \frac{\fs}{\sqrt{2}}  + \sum_{i=1}^3 \frac{\ffi}{\sqrt{2}}\,\moki
- \frac{\fz}{\sqrt{2}}\,\moko\,\mokt\,\mokth - 2\,\fth\,\moa,\,\,\,\,
\label{DGKTsuper}\end{eqnarray}
the flux integers $\fs,\,\ffi,\,\fz,\,\fth$ arising from $F_6,\,F_4,\,F_0,\,F_3$, respectively.
We have turned $F_2$ off, as all results are qualitatively similar, although it is simple to include.

As mentioned, DGKT are able to satisfy the ``tadpole condition" by including an O6-plane.
In order to so, the following relationship between two of the flux integers must hold: 
\begin{equation}
\fz\,\fth = -2.
\label{tadpole}\end{equation}

At this point there are several flux integers in the problem. However, we can simplify the problem
greatly by exploiting the shift and scaling symmetries that we discussed in Section \ref{DiagonalTorus}.
Let us perform the following transformations of our fields:
\begin{eqnarray}
\moki \amp\to\amp \frac{1}{|\ffi|}\sqrt{\frac{|\ffo\,\fft\,\ffth|}{|\fz|}}\,\moki\,\,\,(i=1,2,3),\nonumber\\
\moa  \amp\to\amp \frac{1}{|\fth|}\sqrt{\frac{|\ffo\,\fft\,\ffth|}{|\fz|}}\,\moa + \frac{1}{2\sqrt{2}}\frac{\fs}{\fth}
\label{transfm}\end{eqnarray}
which leaves the form of the kinetic terms invariant. In terms of these new variables, the
superpotential becomes
\begin{eqnarray}
W &=& \sqrt{\frac{|\ffo\,\fft\,\ffth|}{|\fz|}}\nonumber\\
  &\times&\left(\sum_{i=1}^3 \frac{\hffi}{\sqrt{2}}\,\moki 
          -\frac{\hfz}{\sqrt{2}}\,\moko\,\mokt\,\mokth - 2\,\hfth\,\moa \right),
\label{DGKTSuper}
\end{eqnarray}
where the `hat' fluxes are just the signs of the fluxes, e.g., $\hfz\equiv \fz/|\fz|$.
This has a very interesting form:
apart from an overall multiplicative factor, the superpotential is {\em independent of the magnitude of the
fluxes} (although their sign will be important).

We now have all the tools we need; the K\"ahler potential and the superpotential.
Using these, we can compute the 4-dimensional potential $V$ using the supergravity formula (\ref{SUGRA}).
Focusing on the symmetric case, i.e., $\moko=\mokt=\mokth$, we find
\begin{eqnarray}
V \amp = \amp V_{\mbox{\tiny{flux}}} \big{[} 
2(3\,\mokao+2\,\sqrt{2}\,\moaa)^2 - 4\,\sgn\,\mokao^3(3\mokao + 2\,\sqrt{2}\,\moaa)\nonumber\\
\amp+\amp 2\,\mokao^6 
+ 6\,\mokgo^2 + 4\,\moag^2 - 12\,\sgn\,\mokao^2\,\mokgo^2 
+ 6\,\mokao^4\,\mokgo^2 + 6\,\mokao^2\,\mokgo^4\nonumber\\
\amp+\amp 2\,\mokgo^6 - 8\,{\sqrt{2}}\,\mokgo^3\moag    
\big{]}/(32\,\mokgo^3\,\moag^4),
\label{DGKTscaled}\end{eqnarray}
where  
\begin{equation}
V_{\mbox{\tiny{flux}}} \equiv \frac{|\fz|^{5/2}|\fth|^4}{|\ffo \fft \ffth|^{3/2}}
\label{DGKTVflux}\end{equation}
is an overall multiplicative scale that depends on the fluxes.
Note that since $\fz$ and $\fth$ are tightly constrained by the tadpole condition (\ref{tadpole}),
$V_{\mbox{\tiny{flux}}}$ is bounded from above and approaches 0 as $\ffo \fft \ffth\to\infty$.
Also, $\sgn\equiv \hfz\hffo\hfft\hffth=\pm 1$, delineates two independent families of $V$. 
The more general result, without simplifying to the symmetrical case, 
is given in Appendix A eq.~(\ref{DGKTfull}).

Here we make a parenthetical comment:
One can perform direct dimensional reduction from the 10-dimensional action without
using the K\"ahler potential or superpotential. In the DGKT paper this is done explicitly
with the axions ($\mokai,\moaa$) set to their SUSY values. 
Having set the axions to their SUSY values, the natural (and perhaps most intuitive) 
co-ordinates are then the original fields: dilaton $\phi$ and radions $\mokgi$. They find
\begin{eqnarray}
V&=&\frac{\fth^2}{4}\frac{e^{2\phi}}{\mbox{Vol}^2}+\frac{1}{4}\left(\sum_{i=1}^3 \ffi^2\,\mokgi^2\right)\frac{e^{4\phi}}{\mbox{Vol}^3} \nonumber\\
&&+\frac{\fz^2}{4}\frac{e^{4\phi}}{\mbox{Vol}}- 2 \frac{e^{3\phi}}
{\mbox{Vol}^{3/2}},
\end{eqnarray}
where the first 3 terms come from fluxes: 3-form, 4-form, and 0-form, respectively,  and the final
term comes from the O6-plane.
The first 3 terms take on the form we indicated in eq.~(\ref{energy}) for $p$-forms. The final
term carries a minus sign, since O6-planes carry {\em negative} tension, as
we indicated in eq.~(\ref{Oplane}). This term is crucial to achieve stability.
In this form it is not clear that the fluxes scale out, however. By rewriting this in terms of the variables
$\mokgi,\,\moag = e^{-\phi}\sqrt{\mbox{Vol}}/\sqrt{2}$, and then scaling according to eq.~(\ref{transfm}),
it is simple to show that one recovers a simpler version of (\ref{DGKTscaled}), one with $\mokai$ and $\moaa$ set to zero. 

\begin{figure}[t]
\begin{center}
\includegraphics[width=0.9\columnwidth]{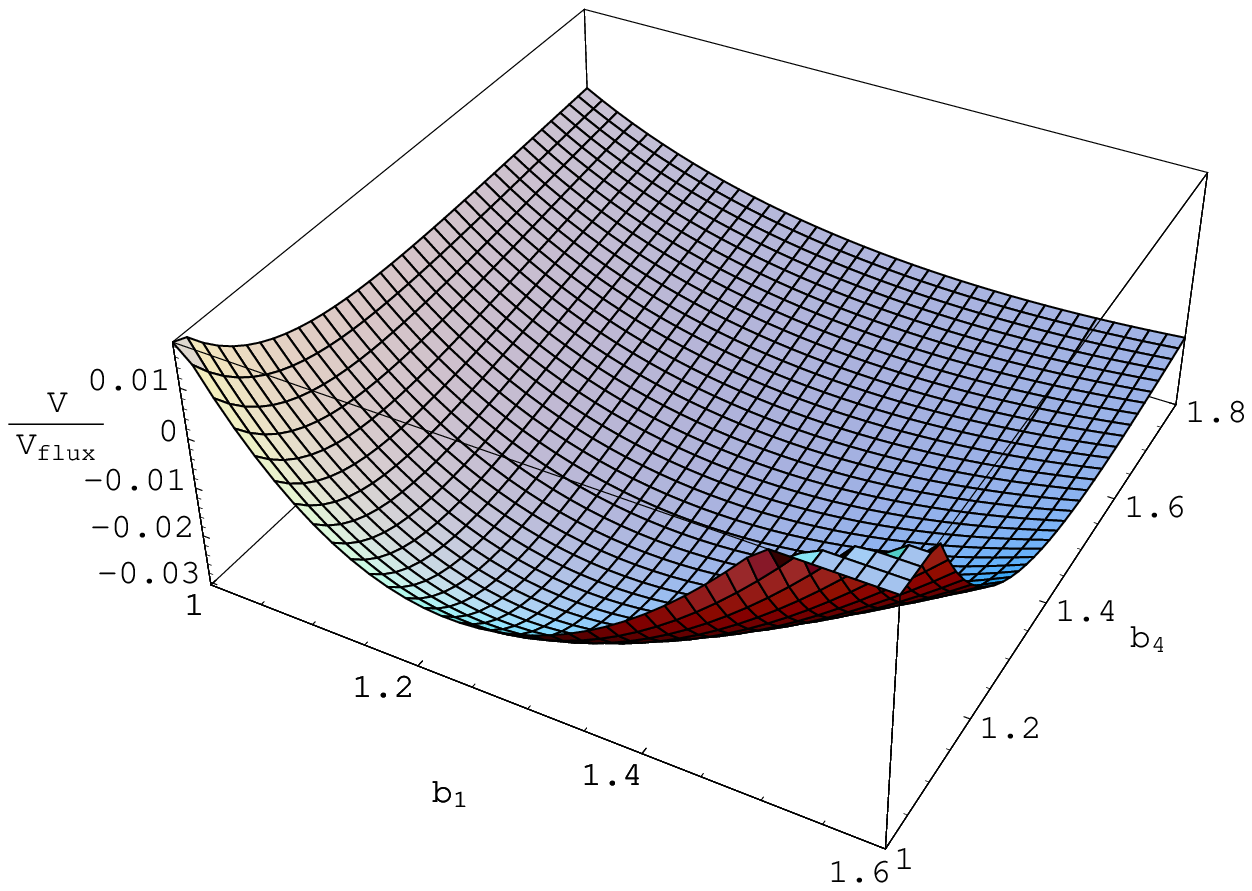}
\includegraphics[width=0.9\columnwidth]{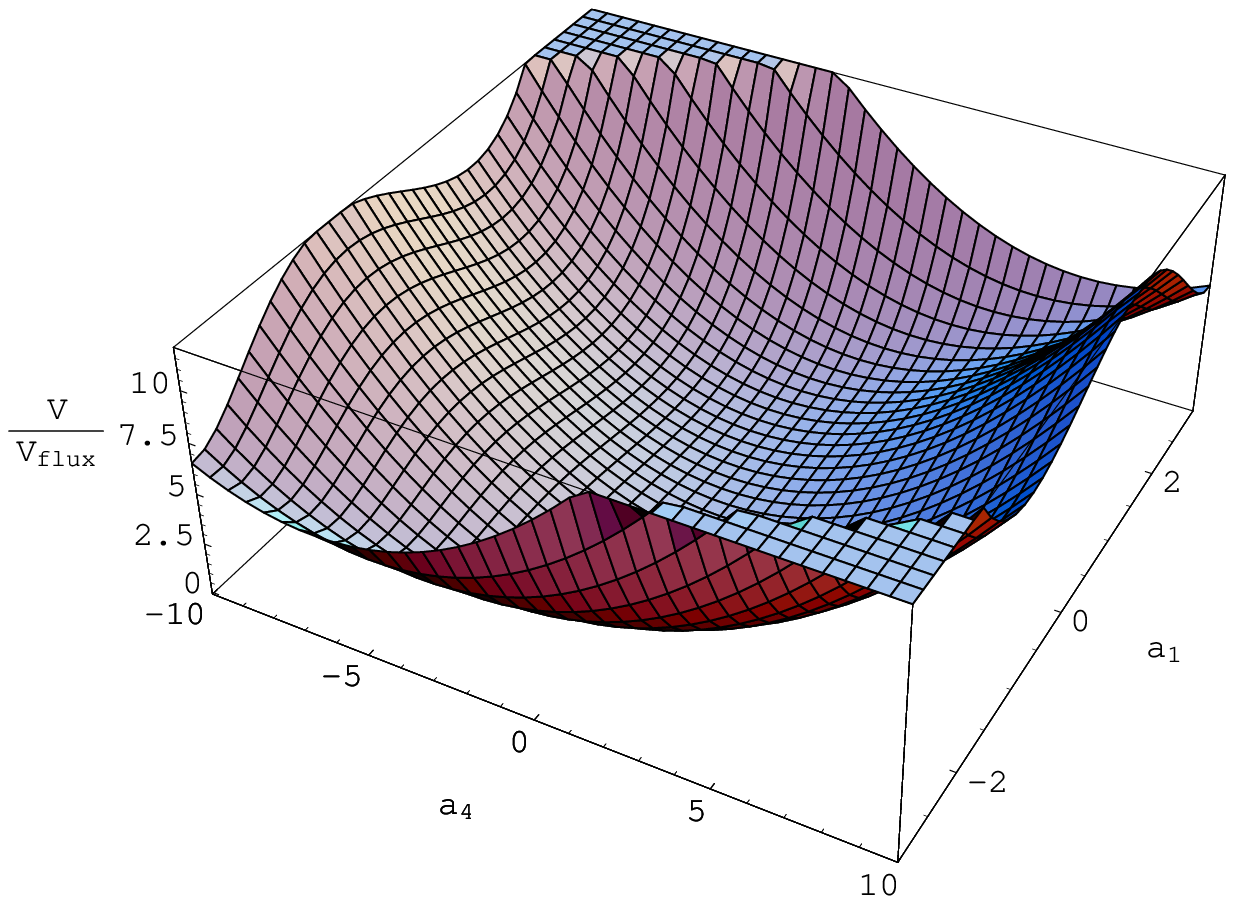}
\caption{Top: The potential $V=V(\mokgo,\moag)$ with axions at their SUSY values: $\mokao=\moaa=0$. 
Bottom: The potential $V=V(\mokao,\moaa)$ with $\sgn=+1$ and geometric moduli at their SUSY values: $\mokgo\approx 1.291,\, \moag\approx 1.217$.}
\label{DGKT_V3D}\end{center}
\end{figure}

We plot $V$ in Fig.~\ref{DGKT_V3D}.
In order to discuss the properties of this potential, let us begin by discussing its supersymmetric properties.
The SUSY vacuum lies at
\begin{equation}
\mokao=0,\,\,\mokgo\approx 1.29,\,\,\moaa=0,\,\,\moag\approx 1.22.
\label{DGKTsusy}\end{equation}
For $\sgn=-1$, this has a corresponding positive definite mass matrix and is clearly stable. However, for
$\sgn=+1$ the mass matrix has negative eigenvalues and so is tachyonic, as reported by DGKT.
Nevertheless, it is stable as it satisfies the Breitenlohner-Freedman bound \cite{Breit}
which states that tachyonic vacua can be stable if the cosmological constant is large and negative.
In Fig.~\ref{DGKT_V3D} (top), we set the axions to zero and plot $V$ as a a function of $\mokgo$ and $\moag$
in the vicinity of the SUSY vacuum. We see that with respect to these two co-ordinates,
the potential has a regular stable minimum. Note also, that with $\mokao=\moaa=0$ then
the values of the potential in (\ref{DGKTscaled}) for $\sgn=\pm 1$ coincide. 
In Fig.~\ref{DGKT_V3D}~(bottom), we plot $V$ with $\mokgo$ and $\moag$ fixed at their SUSY values, 
and allow $\mokao$ and $\moaa$ to vary. We have plotted the case $\sgn=+1$ as its behavior is the most interesting. 

\begin{figure}[t]
\begin{center}
\includegraphics[width=0.9\columnwidth]{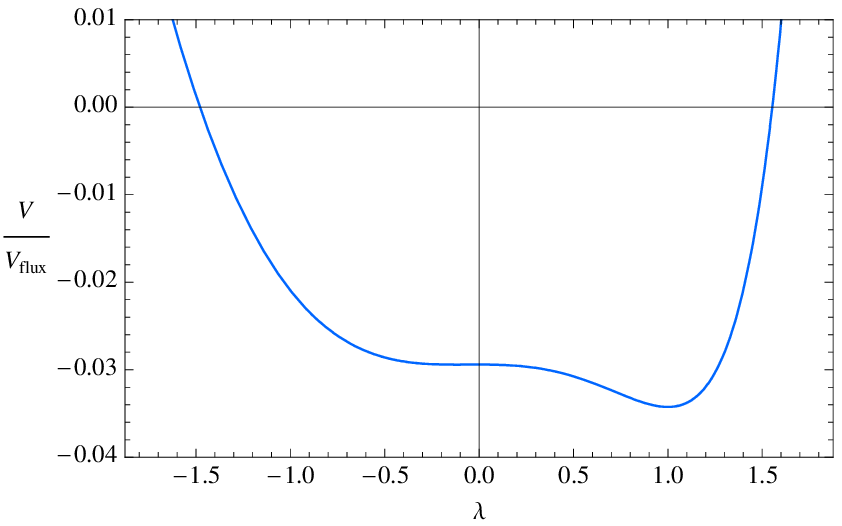}
\includegraphics[width=0.9\columnwidth]{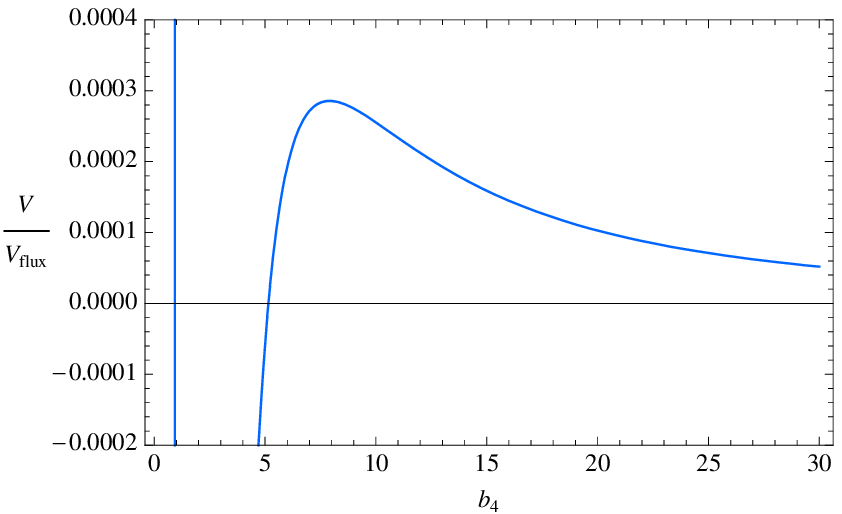}
\caption{Top: We plot $V=V(\lambda)$ by interpolating between the two stationary points of the potential, 
which exists for the `lower case'. $\lambda=0$ corresponds to the (tachyonic) 
SUSY vacuum and $\lambda=1$ corresponds to a local (non-SUSY) minimum.
Bottom: We plot $V=V(\moag)$, focusing on large $\moag$, with all other moduli fixed at their SUSY values.}
\label{DGKT_V2D}\end{center}
\end{figure}

Now, let us investigate the potential further away from the SUSY point. We find that
in the $\sgn=+1$ case (tachyonic), there is a second stationary point of the potential. It is non-supersymmetric, and lies at:
\begin{equation}
\mokao\approx\pm 0.577,\,\,\mokgo\approx 1.15,\,\,\moaa\approx\mp 0.544,\,\,\moag\approx 1.09.
\end{equation}
Given these two stationary points of $V$ (one SUSY, one non-SUSY) we choose to plot $V$ as a function
of $\lambda$, where $\lambda$ is a parameter that linearly interpolates between these two points.
With the SUSY point denoted by a vector of moduli $\gmo_{i,susy}$ and the other (non-SUSY)
stationary point denoted by a vector $\gmo_{i,stat}$, we form the interpolating vector:
\begin{equation}
\gmoi(\lambda)\equiv (1-\lambda) \gmo_{i,susy}+\lambda \gmo_{i,stat}
\end{equation}
so that $\lambda=0$ is the SUSY vacuum and $\lambda=1$ is the second stationary point.
We plot this in Fig.~\ref{DGKT_V2D} (top). 
Also, in Fig.~\ref{DGKT_V2D} (bottom), we plot $V$ as a function of $\moag$ with all other moduli at their SUSY values.
As already stated, $\moag\approx 1.217$ is the SUSY point (a minimum with respect to $\moag$) and this
exists on the left hand side of the figure with the potential much lower than shown. However, the interesting
feature is that for $\moag\approx 7.912$ there exists a local maximum (with $V>0$) with respect to this modulus (but not stationary with respect to the other moduli)
and then $V$ approaches zero from above as $\moag\to\infty$.
There is quite similar behavior when one plots $V$ versus the radial modulus, with the dilaton fixed.

Let us recapitulate the salient features of this class of vacua.
We have come to an important realization: the potential $V$ is of the form $V=V_{\rm flux}(f_i)\,\mbox{func}(\mo^i)$,
where $f_i$ are flux integers and $\mbox{func}(\mo^i)$
is some function of the (rescaled) moduli, {\it independent of fluxes}.
Hence, apart from the overall multiplicative scale (which is proportional to the cosmological constant) 
all vacua look the same. This means that the slow-roll parameters $\epsilon$ and $\eta$ are 
independent of the fluxes. So for this model inflation is realized by all or none of the flux vacua.

Of course we wish to know if the potential is sufficiently flat in some region to exhibit slow-roll.
Here we see a general barrier to this. Note that the potential is a polynomial in 
$\{\gmoai, \gmogi, 1/\gmogi\}$. Naively, this may look as if
it allows for inflation due to some form of power law potential, 
e.g., the potential is quadratic in $\moaa$,
so this may look like Linde's $\sim \phi^2$ ``chaotic inflation" \cite{Chaotic_1,Chaotic_2}.
However, by inspecting the form of eq.~(\ref{epsilon}), we see that this is not at all the case.
The factor of $\gmogi^2$ in the summand changes the picture significantly.
It means that the typical contribution to $\epsilon$ is not $\mathcal{O}(\phi^{-2})$ but $\mathcal{O}(1)$,
and cannot be tuned small by taking $\phi$ large, as in chaotic inflation.
In fact, an extensive numerical search of moduli space (detailed below) suggests that $\epsilon>1$ 
whenever $V>0$ (of course $\epsilon\to 0$ at the stationary point(s) of the potential, but $V<0$ there).
With the axions set to zero, it is simple to analytically prove the non-existence of inflation. With axions non-zero, we
produced vast tables of $\epsilon$ supporting this result.
We will give a representative plot of $\epsilon$ in the upcoming VZ model (see Fig.~\ref{VZ_eps}).

\subsection{The Model of Villadoro and Zwirner (VZ)}\label{VZ}
In March 2005, Villadoro and Zwirner (VZ) \cite{Villadoro} 
constructed a class of orientifold compactifications based on
toroidal orbifolds, where the dilaton and all the moduli associated
with the torus are stabilized  through the inclusion
of $p$-form field strength fluxes (as discussed earlier) and other sorts of 
fluxes, simply referred to as {\em general fluxes}. Their model is
strongly motivated by the work of Derendinger et~al. \cite{Derendinger}.
In particular, VZ include what are known as Scherk-Schwarz geometrical fluxes\footnote{
Geometric flux here refers to a particular kind of topologically non-trivial alteration of the metric on the compact space which yields a contribution to the scalar potential analogous to the contributions from the $p$-form fluxes.}
which provides a large class of vacua.
With many fluxes in the model, there are a number of Bianchi identities and tadpole constraints that the fluxes
must satisfy. This is achieved by including D6-branes and O6-planes. The interested reader
is referred to the original paper \cite{Villadoro} for details.

As originally studied by Derendinger et.~al.~\cite{Derendinger} the orbifold is
$T^6/\mathbb{Z}_2$. A further $\mathbb{Z}_2$ projection is performed to obtain an O6 orientifold. 
This particular orientifold permits 6 degrees of freedom in the metric on the compact space. 
The torus takes the form $T_6=T_2\times T_2\times T_2$ and possesses a diagonal metric.
Then, without loss of generality, the 6-dimensional metric can be parameterized by 
6 variables $\gamma_i$ and $\beta_i$ (i=1,2,3) as follows:
\begin{eqnarray}
(ds^2)_6 \amp =\amp \sum_{i=1}^3\left((\gamma_i/\beta_i)\,(dx^i)^2+(\gamma_i\,\beta_i)\,(dy^i)^2\right), \label{metricVZ}\\
\mbox{Vol} \amp = \amp \int_{T^6/\mathbb{Z}_2^2}
d^6x\sqrt{g_6}=\gamma_1\gamma_2\gamma_2 = \vmokgo\vmokgt\vmokgth.
\end{eqnarray}
The form of the volume explains the choice in decomposing the metric as above, namely that the product
of the $\gamma_i$ is proportional to the volume, as it was for DGKT. In turn, we again identify 3 good
K\"ahler co-ordinates $b_1,\,b_2,\,b_3$. The $\beta_i$, on the other hand, are related to the complex
structure and axio-dilaton moduli.

Let us now provide the full list of moduli (again ignoring ``blow-up modes"). In this case there are seven complex moduli
that survive the orientifold projection: 3 K\"ahler moduli $\vmoki=\vmokai+i\vmokgi,\,\,i=1,2,3$, 
an axio-dilaton $\vmoa=\vmoaa+i\vmoag\,(\vmoag=e^{-\phi}\sqrt{\mbox{Vol}}/\sqrt{\beta_1\,\beta_2\,\beta_3}$), 
and 3 complex structure moduli $\vmoci=\vmocai+i\vmocgi,\,\,i=5,6,7\,\,
(\vmocgo=e^{-\phi}\sqrt{\mbox{Vol}}\sqrt{\beta_2\,\beta_3/\beta_1}$ etc).

The K\"ahler potential takes on an extremely simple form: in the notation of Section \ref{DiagonalTorus}
it has all 7 $n_i=1$. Explicitly, it is\footnote{We follow the convention of Ref.~\cite{Villadoro} where 
an overall factor of $2^7$ was removed from the argument of the logarithm,
since it can be simply reabsorbed into $W$.}
\begin{equation}
K = -\ln\left(\vmokgo\,\vmokgt\,\vmokgth\,\vmoag\,\vmocgo\,\vmocgt\,\vmocgth\right)
\label{VZKahler}\end{equation}
By way of comparison to the DGKT model, it is as though 
$\vmoag^4\to\vmoag\,\vmocgo\,\vmocgt\,\vmocgth$,
in order to accommodate the 3 complex structure moduli that appear here.

The superpotential incorporates geometric flux, in addition to the familiar $p$-form flux, 
and satisfies the `tadpole condition' with D-branes. 
The superpotential, as derived in \cite{Derendinger}, is given by:
\begin{eqnarray}
W \amp = \amp \lamooo-\lamoot(\vmoko+\vmokt+\vmokth)
+\lamttt\,\vmoko\,\vmokt\,\vmokth\nonumber\\
\amp + \amp \lamott(\vmoko\,\vmokt+\vmoko\,\vmokth+\vmokt\,\vmokth) 
- \lamooop\,\vmoa \nonumber\\
 \amp + \amp \lamootp\,\vmoa(\vmoko+\vmokt+\vmokth)-\lamoof(\vmoco+\vmoct+\vmocth)\nonumber\\
\amp + \amp \lamotf(\vmoko(\vmoct+\vmocth)+ \vmokt(\vmoco+\vmocth)+\vmokth(\vmoco+\vmoct))
\nonumber\\
\amp - \amp \lamooth(\vmoko\,\vmoco+\vmokt\,\vmoct+\vmokth\,\vmocth)
\end{eqnarray}
Here we have designated the fluxes by $f_{ijk}$, which correspond to different choices of $p$-form and
geometric flux wrapped on various cycles of the torus.
There are constraints that the fluxes $f_{ijk}$ must satisfy, namely
\begin{equation}
\lamotf(\lamotf-\lamooth)=0,\,\,\,\lamootp(\lamotf-\lamooth)=0.
\label{constraints}\end{equation}

VZ find a family of SUSY vacua by choosing the following parameterization of fluxes:
\begin{eqnarray}
\amp\amp\lamooo=-15 \None,\,\,\lamoot=\frac{3\Ntwo}{\tO},\,\,\lamott=\frac{\None}{\tO^2},\nonumber\\
\amp\amp\lamttt=-\frac{3\Ntwo}{\tO^3},\,\,\lamooop=-\frac{2\Ntwo}{\sO},\,\,\lamootp=-\frac{2\None}{\tO\sO},\nonumber\\
\amp\amp\lamoof=-\frac{2\Ntwo}{\uO},\,\,\lamooth=\lamotf=-\frac{6\None}{\tO \uO}.
\end{eqnarray}
We note that the $f_{ijk}$ are actually non-integer. 
Here we do not record the conditions that $\None,\,\Ntwo,\,\tO,\,\sO,\,\uO$ must satisfy, but refer the reader to \cite{Villadoro}. 
We do note that $\{\tO,\,\sO,\,\uO\} \in \mathbb{R}^{+}$.
What is important is that this designates an infinite family of vacua
with fluxes parameterized by the set of five parameters $\None,\,\Ntwo,\,\tO,\,\sO,\,\uO$. 
So we have started with
a superpotential with 9 fluxes: $\lamooo,\ldots,\lamotf$,
one has been eliminated by the conditions (\ref{constraints}) ($\lamooth=\lamotf$),
three have been eliminated by demanding that the SUSY condition (\ref{SUSY}) be 
satisfied for each $\vmoki$, leaving five independent parameters.
 
 For this family of vacua it is rather straightforward to show that we can scale out
 the fluxes $\tO,\,\sO,\,\uO$ by making the following rescaling of our fields:
\begin{eqnarray} 
&&\vmoki\to \tO\,\vmoki\,\,(i=1,2,3),\,\,\,\vmoa\to \sO\,\vmoa,\nonumber\\
&&\vmoci\to \uO\,\vmoci\,\,(i=5,6,7).
\end{eqnarray} 
This leaves only $\None$ and $\Ntwo$ of which we can scale out one of them, leaving only their
ratio as a tunable parameter 
\begin{equation}
\rat\equiv \frac{\None}{\Ntwo}
\end{equation}
($\Ntwo=0$ can be handled separately).

Now let us focus on the symmetric case, in which $\vmoko=\vmokt=\vmokth$ and $\vmoco=\vmoct=\vmocth$, and keep track of the fields $\vmoko,\,\vmoa,\,\vmoco$.
We find that $W$ is simplified to
\begin{eqnarray}
W\amp=\amp-15\None - 9\Ntwo\vmoko + 3\None\vmoko^2 - 3\Ntwo\vmoko^3 \nonumber\\ \amp\amp 
+ 2\Ntwo(\vmoa + 3 \vmoco) - 6\None\vmoko(\vmoa + 3 \vmoco) .
\label{VZSuper}\end{eqnarray}
We note that since $W$ only depends on a linear combination of $\vmoa$ and $\vmoco$,
namely $\vmoa + 3 \vmoco$, the potential $V$ only depends on the same linear
combination of the corresponding axions, namely $\vmoaah\equiv \vmoaa+3\vmocao$.

By using (\ref{SUGRA}), it is a straightforward matter to obtain the potential. 
The result is a rather long expression that we report in Appendix B eq.~(\ref{VZfull}).
The leading prefactor
\begin{equation}
V_{\mbox{\tiny{flux}}} \equiv \frac{\Ntwo^2}{\tO^3\,\sO\,\uO^3}
\label{VZVflux}\end{equation}
is an overall multiplicative scale that depends on the fluxes. 
At fixed $\rat$  there exists a family of solutions for
$\None,\Ntwo,\,\tO,\,\sO,\,\uO$ for which $V_{\mbox{\tiny{flux}}}\to 0$ parametrically.
However, one should note the explicit appearance of $\rat=\None/\Ntwo$ in the potential,
which controls its {\em shape}.
We mention that without loss of generality we can focus on $\rat$ non-negative, since
$\rat\to-\rat$ and $\vmokai\to-\vmokai$ leaves $V$ unchanged.

By solving the equations $D_iW=0$, one can show that the SUSY vacuum lies at:
\begin{equation}
\vmokao =\vmoaa+3\,\vmocao = 0,\,\,\,\vmokgo = \vmoag = \vmocgo = \sqrt{\frac{5}{3}}\,,
\end{equation}
for all  $\rat$.
We note that this (AdS) SUSY vacuum is tachyonic but stable, as is satisfies the Breitenlohner-Freedman bound \cite{Breit}. 
The potential here takes on the value
\begin{equation}
V_{\mbox{\tiny{SUSY}}} = - V_{\mbox{\tiny{flux}}} \frac{432\sqrt{15}}{125} \left(1+15\rat^2\right).
\end{equation}
Now, an important special case is when $\None=\Ntwo$ ($\rat=1$), since as VZ describe, this provides
this $\mathcal{N}=1$ supergravity theory with an interpretation in terms of an 
$\mathcal{N}=4$ supergravity theory with extended (gauged) symmetry. This may
be of some interest \cite{Derendinger2}. In this case, the potential in eq.~(\ref{VZfull}) may be simplified to
\begin{eqnarray}
V \amp = \amp V_{\mbox{\tiny{flux}}}(
36 \tilde{a}_1^6+108 b_1^2 \tilde{a}_1^4+144 \tilde{a}_4 \tilde{a}_1^4+1280 \tilde{a}_1^3/3 \nonumber\\
\amp+\amp 108 b_1^4 \tilde{a}_1^2+144 \tilde{a}_4^2 \tilde{a}_1^2+144 \tilde{a}_4 b_1^2 \tilde{a}_1^2+144 b_4^2 \tilde{a}_1^2\nonumber\\
\amp+\amp 432 b_5^2 \tilde{a}_1^2+2560 \tilde{a}_4 \tilde{a}_1/3
+36 b_1^6+48 \tilde{a}_4^2 b_1^2+48 b_1^2 b_4^2 \nonumber\\
\amp - \amp 432 b_1^2 b_5^2-576 b_1^2 b_4 b_5+102400/81)/(b_1^3\,b_4\,b_5^3).
\end{eqnarray}
with $\tilde{a}_1\equiv a_1-1/3$ and $\tilde{a}_4=a_4+3 a_5+4/3$.
In addition to the SUSY vacua, we find three additional AdS vacua given by
\begin{eqnarray}
\vmokao \amp=\amp\vmoaa+3\vmocao = \frac{1}{3},\,\,\vmokgo \approx 1.38,\,\,\vmoag = \vmocgo \approx 1.26\nonumber\\
\vmokao \amp=\amp\vmoaa+3\vmocao = \frac{1}{3},\,\,\vmokgo \approx 1.38,
\,\,\vmoag \approx 2.87,\,\,
\vmocgo \approx 0.958\nonumber\\
\vmokao \amp=\amp 1,\,\vmoaa+3\vmocao = -4,\,\,\vmokgo = \vmoag = \vmocgo = \frac{2}{\sqrt{3}}
\label{AdSeq}\end{eqnarray}

\begin{figure}[t]
\begin{center}
\includegraphics[width=0.9\columnwidth]{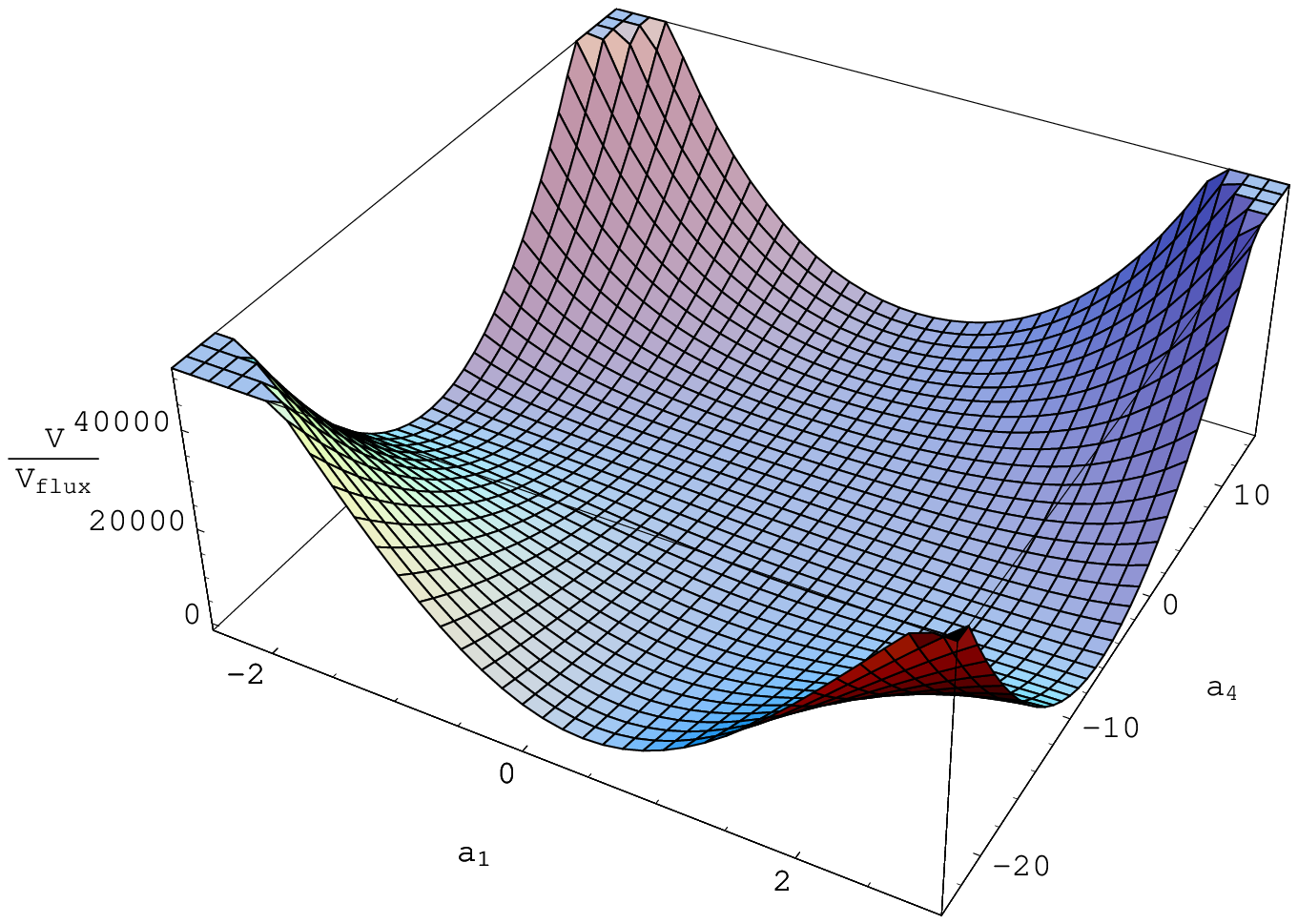}
\includegraphics[width=0.9\columnwidth]{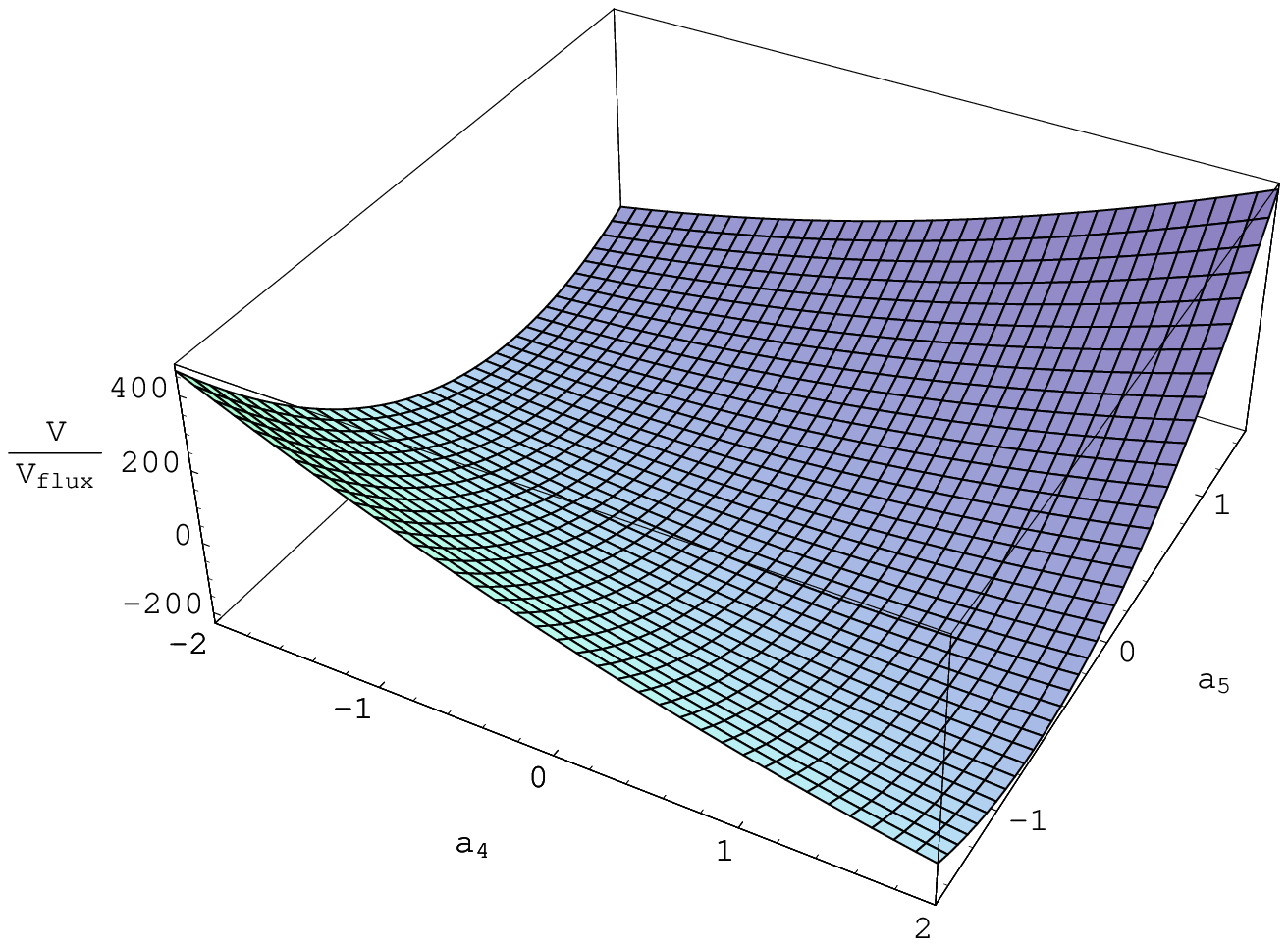}
\caption{Top: The potential $V=V(\vmokao,\vmoaa)$ with $\vmocao=0$ and 
other moduli taking on their SUSY values. 
Bottom: The potential $V=V(\vmoaa,\vmocao)$ with
other moduli taking on their SUSY values. }
\label{VZ_V}\end{center}
\end{figure} 
In Fig.~\ref{VZ_V} (top) we plot $V$ as a function of $\vmokao$ and $\vmoaa$ with other parameters fixed at their SUSY values.
We see that $V$ is relatively flat along each axis, while $V$ is steep along diagonals.

Now, this potential contains one modulus that is not stabilized. One linear combination
of the axions is left exactly massless at the SUSY vacuum. We we will return to this later in the discussion. 
This is a result of the fact that the superpotential in eq.~(\ref{VZSuper}) only depends
on the combination $\vmoaa+3\,\vmocao$.
A plot of $V$ as a function of $\vmoaa$ and $\vmocao$,
with all other moduli fixed at their SUSY values, is given in Fig.~\ref{VZ_V} (bottom). We see the flat `valley'.
The existence of such a flat direction certainly seems useful from the point of view of inflation,
however
one should recall that this flat direction emanates from an AdS vacuum.

\begin{figure}[t]
\begin{center}
\includegraphics[width=0.9\columnwidth]{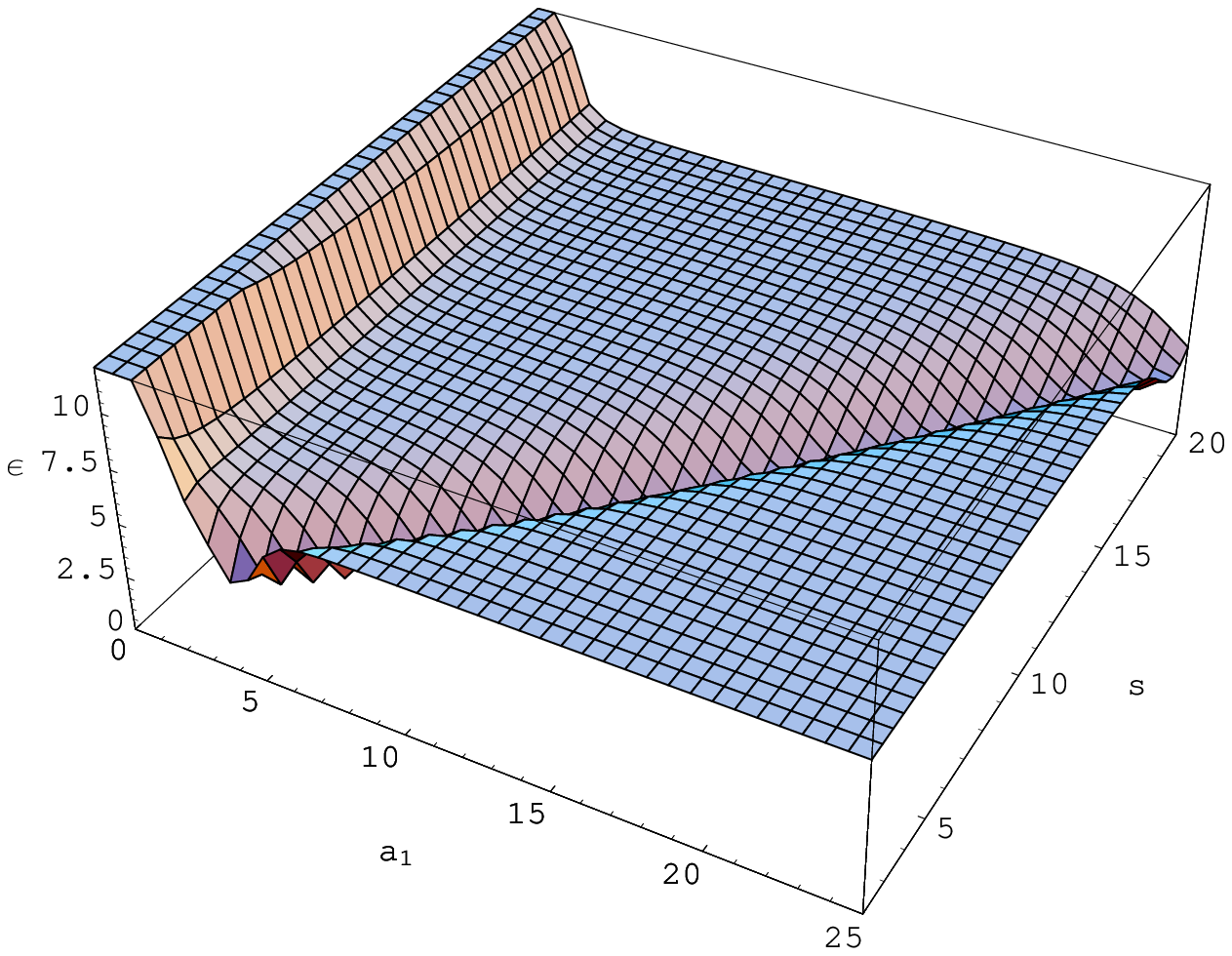}
\includegraphics[width=0.9\columnwidth]{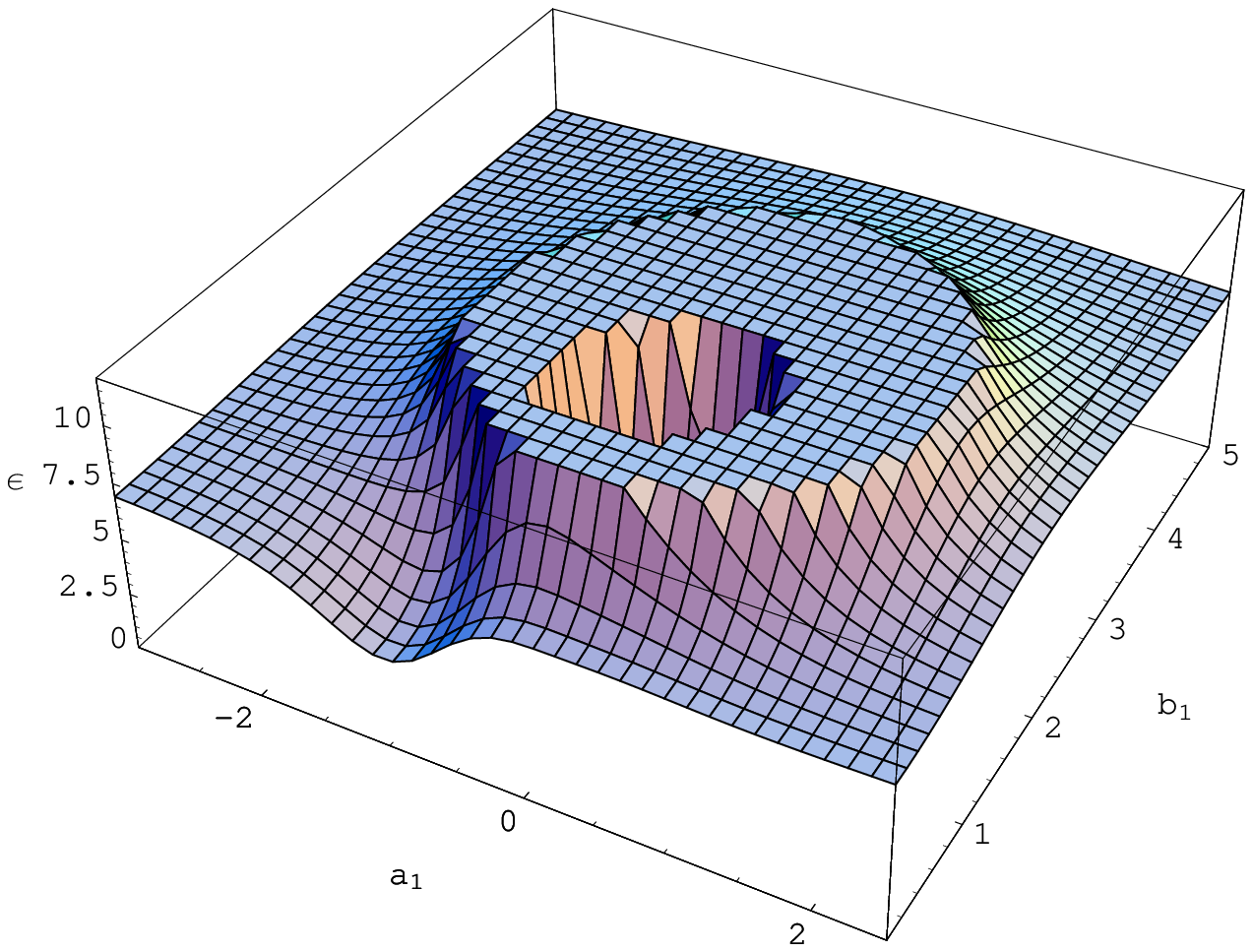}
\caption{
Top: The slow-roll parameter $\epsilon(\vmokao,\rat)$ with other moduli taking on their values
from the third set of eq.~(\ref{AdSeq}).
Bottom: The slow-roll parameter $\epsilon(\vmokao,\vmokgo)$ with $\rat=1$ and other moduli taking on 
their SUSY values.}
\label{VZ_eps}\end{center}
\end{figure}

For slow-roll it is again evident that this is very difficult, due to the argument presented at
the end of Section \ref{DGKT}, namely that the characteristic value of $\epsilon$ in these types
of tree-level toroidal models is $\mathcal{O}(1)$. However, it is important to investigate
the effect of our tunable parameter $\rat$. To get a flavor of its effect, in
Fig.~\ref{VZ_eps} (top) we plot $\epsilon=\epsilon(\vmokao)$ for
$2<\rat<20$. We see that $\epsilon>1$ in this region. Indeed our numerical
studies indicate (detailed below) that there is no inflating region anywhere in moduli space.
Again this is based on the results of vast tables of $\epsilon$ over moduli space.
For a more conventional representation,
in Fig.~\ref{VZ_eps} (bottom) we plot $\epsilon$ as a function of a pair of moduli, namely $\vmokao$ and $\vmokgo$, with $\rat=1$. Here $\epsilon$ is large in all regions in which $V>0$. ($\epsilon\to\infty$ as $V\to 0$ and $\epsilon\to 0$ at the AdS minimum).
The plot displays a dip in $\epsilon$ as $\vmokao\to-1,\,\vmokgo\to 0$. 
At this point we find $\epsilon\to 4$.

\subsection{The Model of Ihl and Wrase (IW)}\label{IW}
In the previous two models, the K\"ahler potential took on the form of eq.~(\ref{Kprod}), which
we referred to as diagonal torus models. It followed from this that the
first slow-roll parameter took on the form as given in eq.~(\ref{epsilon}). For potentials $V$ that 
were rational in the moduli, this meant that $\epsilon=\mathcal{O}(1)$ was quite natural.
We would like then to investigate more complicated models in which this does not occur. 
 In April 2006 Ihl and Wrase \cite{Ihl} obtained an explicit example of this nature. Their work is strongly motivated by the work of DGKT. Indeed they also consider massive type IIA supergravity. 
However, unlike the DGKT model, we find that one tunable parameter remains in the
potential, as we found in the VZ model.

The orientifold is $T^6/\mathbb{Z}_4$. Unlike the torii of DGKT and VZ, this orientifold
{\em does not} permit the decomposition of $T^6$ to $T_2\times T_2\times T_2$ with identical $T_2$s.
Instead the $T_2$s must have {\em different} metrics. 
The interested reader is referred to the original paper \cite{Ihl} for details. 

The metric on the compact space is somewhat more complicated than our previous models.
When expressed in terms of the most useful co-ordinates (those that are readily related to
K\"ahler and complex structure moduli) the metric on the compact space is non-diagonal.
Here there are 4 independent degrees of freedom that appear explicitly in the metric. Denoting them as usual by
$\gamma_i$, the metric is given by
\begin{eqnarray}
(ds^2)_6 \amp =\amp \sum_{i=1}^3\gamma_i \left[(dx^i)^2+(dy^i)^2\right]\nonumber\\
\amp+\amp 2\,\gamma_4\!\left(\!dx^1\,dx^2+dy^1\,dy^2 - \sum_{i,j=1}^2\epsilon_{ij}\,dx^i\,dy^j\!\right),
\,\,\,\,\,\,\,\,\, \\
\mbox{Vol}\amp = \amp \int_{T^6/\mathbb{Z}_4}d^6x\sqrt{g_6}=U_2\gamma_3(\gamma_1\gamma_2-2\gamma_4^2)/4 \nonumber\\
\amp = \amp \imokgth(\imokgo\imokgt-\imokgf^2/2), \label{VolIW}
\end{eqnarray}
where $\epsilon_{ij}$ is the Levi-Civita symbol defined by 
$\epsilon_{12}=-\epsilon_{21}=1$, $\epsilon_{11}=\epsilon_{22}=0$.
The form of the volume requires a little explanation: The square-root of the 
determinant of the metric is easily shown to be
$\gamma_3(\gamma_1\gamma_2-2\gamma_4^2)$, and indeed $\mbox{Vol}$ is proportional to this.
There is also a factor of $U_2$, which is related to the canonical complex structure moduli.
It is known as a `pure type' contribution that is in some sense hidden in the metric. 
The interested reader is referred to footnote 11 of the IW paper for clarification. 
Again we have written the volume in terms of some $\imokgi$ that are good K\"ahler 
co-ordinates.\footnote{In the IW paper: $\mbox{Vol}=\kappa\,\imokgth(\imokgo\imokgt-\imokgf^2/2) $. By rescaling $\mokg_i\to\kappa^{-1/3}\mokg_i,\,i=1,2,3,4$ we obtain eq.~(\ref{VolIW}) and $\kappa$ is eliminated.}
Note that the volume is {\em not} simply a product of the K\"ahler co-ordinates.

In summary, after the orientifolding there remains \emph{one} complex structure modulus $U_2$. In total we have
4 K\"ahler moduli $\moki=\mokai+i\mokgi,\,i=1,\ldots,4$ and 2 other moduli which mix the axio-dilaton 
and the complex structure modulus: $\imoa=\imoaa+i\imoag \,(\imoag=e^{-\phi}\sqrt{\mbox{Vol}}/\sqrt{U_2})$
and $\imoc=\imoca+i\imocg \,(\imocg=2\sqrt{U_2}\,e^{-\phi}\sqrt{\mbox{Vol}})$.\footnote{The canonical axions are $\xi_5 = 2 \imoaa$ and $\xi_6 = 2 \imoca$.} (We again ignore the ``blow-up modes").

Here the K\"ahler potential {\em does not} take the form of the previous 2 models, i.e.,
it is not of the form of a logarithm of a product of geometric moduli
and so is not quite of the form discussed in Section \ref{DiagonalTorus}. There is a modification due to the
non-trivial form of the volume, namely $\mbox{Vol}=\imokgth(\imokgo\imokgt-\imokgf^2/2)$.
The K\"ahler potential is
\begin{equation}
K = -\ln\left(2\,\imokgth(\imokgo\,\imokgt-\imokgf^2/2) \imoag^2\,\imocg^2\right)
\label{IWKahler}\end{equation}
--- see eq.~(\ref{Kprod}) for comparison.

The ingredients for the superpotential are just the same
as the DGKT model. A 0-form, 3-form, 4-form (and an unimportant 2-form) are included.
The superpotential is a simple modification of the DGKT model, namely
\begin{eqnarray}
W = \amp\amp\frac{\fs}{\sqrt{2}}  + \sum_{i=1}^4 \frac{\ffi}{\sqrt{2}}\,\moki \nonumber\\
\amp \amp - \frac{\fz}{\sqrt{2}}\,\imokth(\imoko\,\imokt-\imokf^2/2) - 2\,\fth(\imoa + \imoc),
\label{IWsuper}\end{eqnarray}
which is to be compared to eq.~(\ref{DGKTsuper}). We note that there is one additional flux component: $\fff$,
which is due to the presence of a 4th K\"ahler modulus $\imokf$.
 
Furthermore, just as in the DGKT model, an $06$-plane is introduced in order to satisfy the
tadpole condition. This occurs in precisely the same way as before (see eq.~(\ref{tadpole})), i.e., $\fz\,\fth = -2$.

Again let us exploit all existing shift and scaling symmetries.
We perform the following transformations on our fields:
\begin{eqnarray}
\moki \amp \to \amp \frac{1}{|\ffi|}\sqrt{\frac{|\ffo\,\fft\,\ffth|}{|\fz|}}\,\moki\,\,\,(i=1,2,3),\nonumber\\
\imokf \amp \to \amp \sqrt{{\frac{|\ffth|}{|\fz|}}}\,\imokf, \nonumber\\
\imoa \amp \to \amp \frac{1}{|\fth|}\sqrt{\frac{|\ffo\,\fft\,\ffth|}{|\fz|}}\,\imoa + \frac{1}{2\sqrt{2}}\frac{\fs}{\fth},
\nonumber\\ \imoc \amp \to \amp \frac{1}{|\fth|}\sqrt{\frac{|\ffo\,\fft\,\ffth|}{|\fz|}}\,\imoc,
\label{Itransfm}\end{eqnarray}
which was chosen in such a way as to leave the kinetic terms invariant.
This allows one to rewrite the superpotential as
\begin{eqnarray}
W = \amp\amp\sqrt{\frac{|\ffo\,\fft\,\ffth|}{|\fz|}}\Big{(}
\sum_{i=1}^3 \frac{\hffi}{\sqrt{2}}\,\moki + \frac{\hfff}{\sqrt{2}}\,\flx\,\imokf \nonumber \\
\amp\amp- \frac{\hfz}{\sqrt{2}}\,\mokth(\imoko\,\imokt-\imokf^2/2)  - 2\,\hfth(\imoa + \imoc) \Big{)}.\,\,\,\,\,\,
\label{IWSuper}\end{eqnarray}
Here there exists one combination of the fluxes that does \emph{not} scale out:
\begin{equation}
\flx\equiv \frac{|\fff|}{\sqrt{|\ffo\,\fft|}}.
\label{IWflx}\end{equation}
This is a result of the non-trivial (``intersection") form for the volume, which puts $\imokf$ on a different
footing from the other K\"ahler moduli.

We turn now to the 4-dimensional potential $V$. 
In the presence of all axions, the result is somewhat complicated: see Appendix C eq.~(\ref{IWfull}).
Here we note that a consistent solution is found with all (shifted) axions vanishing, and so we will focus on this case:
$\imokao=\ldots=\imoca=0$.
Also note that $\imoa$ and $\imoc$ are treated on equal footing. Inspired by this fact, we will concentrate on the case $\imoa=\imoc$. 
This sets $U_2=1/2$, as reported in the IW paper. We also make a final set of multiplicative transformations by $\pm 1$, namely:
$\imokgo\to\hffo\,\imokgo,\,\imokgt\to\,\hffo\,\imokgt,\,\imokgf\to\hfff\,\imokgf$, 
which \emph{still} preserves the form of the kinetic energy.
We find
\begin{eqnarray}
V \amp = \amp V_{\mbox{\tiny{flux}}}[
2(\imokgo^2+\imokgt^2+\imokgth^2)+2\,\sgnot\,\imokgf^2+16\,\imoag^2 \nonumber\\
\amp - \amp 16\,\sqrt{2}\,\imokgo\,\imokgt\,\imokgth\,\imoag + 8\,\sqrt{2}\,\imokgth\,\imokgf^2\,\imoag 
 + 2\,\imokgo^2\,\imokgt^2\,\imokgth^2 \nonumber\\
\amp-\amp 2\,\imokgo\,\imokgt\,\imokgth^2\,\imokgf^2 + \imokgth^2\,\imokgf^4/2
+4(\imokgo+\sgnot\,\imokgt)\imokgf\,\flx \nonumber\\ \amp+\amp(\imokgf^2+2\,\imokgo\,\imokgt)\,\flx^2
]/(2\,\imokgth(\imokgo\,\imokgt-\imokgf^2/2) \imoag^4),\,\,\,\,\,\,\,\,\,
\label{IWpot}\end{eqnarray}
where
\begin{equation}
V_{\mbox{\tiny{flux}}} \equiv \frac{|\fz|^{5/2}|\fth|^4}{|\ffo\,\fft\,\ffth|^{3/2}}
\label{IWVflux}\end{equation}
as we defined it for the DGKT model. Note however, that this is not the only piece that depends on the magnitude
of the fluxes, since the flux parameter $\flx$ also appears in (\ref{IWpot}). 
So there is one combination of the fluxes that describes the {\em shape} of the potential. 
We have defined $\sgnot\equiv\hffo\hfft=\pm 1$ which delineates two families of potentials.
We should also keep track of the physical constraints
that the area of the third torus and the compact volume $\mbox{Vol}$ are both positive, so $\imokgth>0$ and $\imokgo\imokgt-\imokgf^2/2>0$.

All stationary points are AdS (even the non-SUSY ones). 
The interested reader is referred to the IW paper \cite{Ihl} for a detailed description of the locations 
of these stationary points.
Here we begin by noting that when $\sgnot=+1$ and $\flx=0$, there is a SUSY AdS minimum, which coincides with that of the DGKT model:
\begin{equation}
\imokgo=\imokgt=\imokgth\approx 1.29,\,\,\imokgf=0,\,\,\imoag\approx 0.609.
\label{IWsusy}\end{equation}
(compare to eq.~(\ref{DGKTsusy}) with renaming of variables; $\imoag$ of IW replaced by $\moag/2$ of DGKT.)

Since the K\"ahler potential is not simply the logarithm of a product of moduli, $\epsilon$ is {\em not} given by eq.~(\ref{epsilon}). 
Instead we revert to eq.~(\ref{SReps}).
We emphasize that the particular transformations we have performed on the $\imoki$ have left
the form of $K^{i\bar{j}}V_i V_{\bar{j}}$ unchanged. With this in mind, we find the following:
\begin{eqnarray}
\epsilon = \frac{1}{V^2}\Bigg{\{}\!\!\!\!\!\!
\amp\amp\sum_{i,j=1}^4 M_{ij}\left(\frac{\partial V}{\partial\imokai}\frac{\partial V}{\partial\imokaj}
 +\frac{\partial V}{\partial\imokgi}\frac{\partial V}{\partial\imokgj}\right) \nonumber\\
\amp+\amp \sum_{i=5}^6\frac{\imokgi^2}{2}\left[\left(\frac{\partial V}{\partial\imokai}\right)^2
 +\left(\frac{\partial V}{\partial\imokgi}\right)^2\right]\Bigg{\}},\,\,\,
\label{IWeps} \end{eqnarray}
where 
\begin{equation}
M \equiv \left(\begin{array}{cccc}
\imokgo^2 & \imokgf^2/2 & 0 & \imokgo\imokgf \\
\imokgf^2/2 & \imokgt^2 & 0 & \imokgt\imokgf \\
0 & 0 & \imokgth^2 & 0 \\
\imokgo\imokgf & \imokgt\imokgf & 0 & \imokgo\imokgt+\imokgf^2/2
\end{array} \right)\,,
\label{IWepsmatrix}\end{equation}
and $V$ given in terms of our rescaled variables, i.e., by eq.~(\ref{IWpot}) for the simple vanishing axion case, 
and by eq.~(\ref{IWfull}) for the general non-vanishing axion case.
\begin{figure}[t]
\begin{center}
\includegraphics[width=0.9\columnwidth]{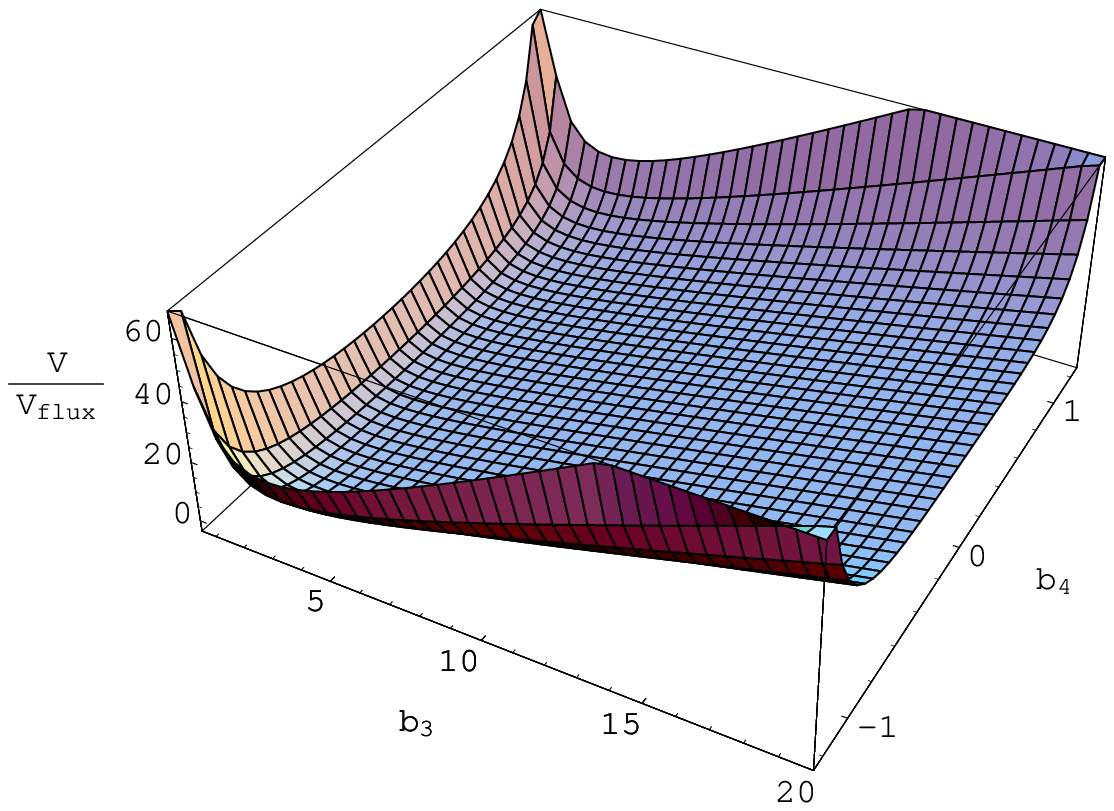}
\includegraphics[width=0.9\columnwidth]{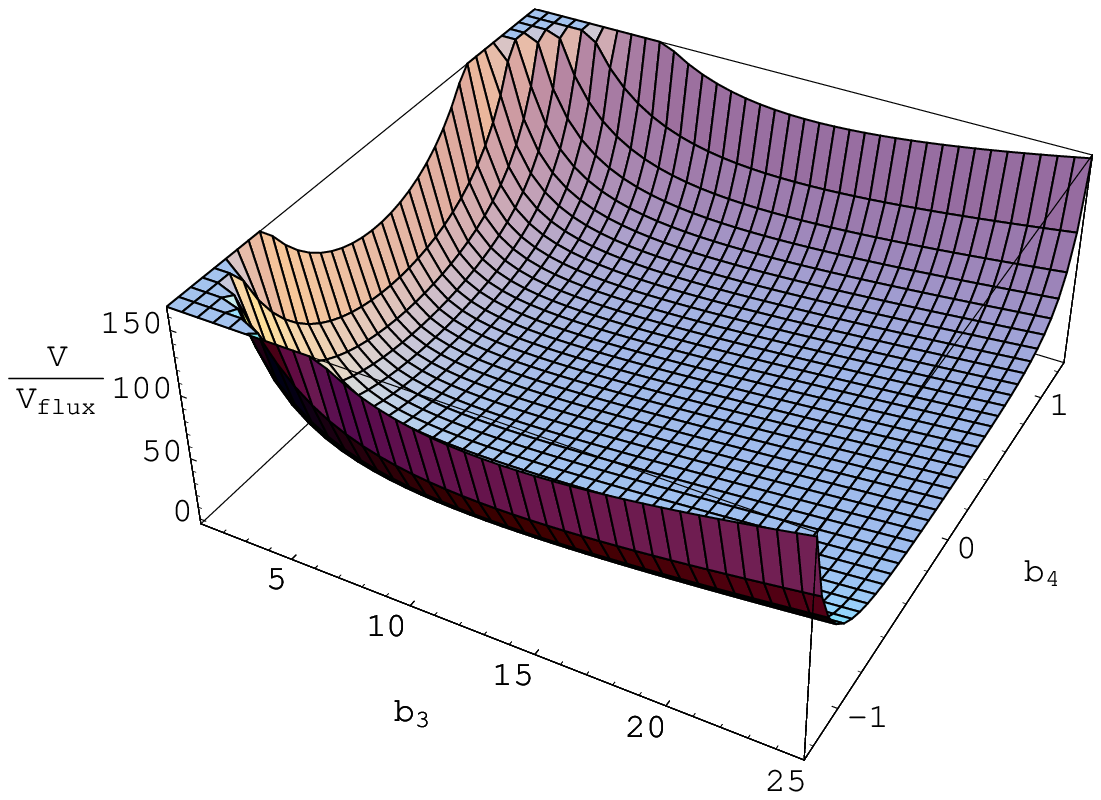}
\caption{The potential $V=V(\imokgth,\imokgf)$ with 
$\imokgo=\imokgt=\imoag=1$ and $\sgnot=-1$.
Top: $\flx=1$. 
Bottom: $\flx=10$.}
\label{IWVplot}\end{center}
\end{figure}
In Fig.~\ref{IWVplot} we give a representative plot of a piece of moduli space. We
see significant variation as we change the flux parameter from $\flx=1$ in (top) to $\flx=10$ in (bottom).
In this plot we have ensured that $\imokgf$ has remained in the physical region given by
$\imokgo\imokgt-\imokgf^2/2>0$. We note that the potential becomes singular at this boundary;
this follows again from the non-trivial form of Vol. Our numerical investigations
into the slow-roll parameter $\epsilon$ have again yielded $\epsilon>1$ whenever $V>0$,
although we have not investigated the full moduli space -- we did
not include all axions in our search.




\subsection{Comments on Blow-up Modes}\label{bu}
In the three models that we have investigated, we have ignored a class of moduli known as ``twisted moduli" or ``blow-up modes". Recall that apart from the dilaton, the geometric moduli describe the size and shape of the compact space, i.e., its {\em geometry}. These are the K\"ahler and complex structure moduli. For a smooth compact space, this is fully general. However, the models investigated here are not smooth; they are all {\em orbifolds}, which have fixed points. These fixed points correspond to conical singularities. In the large volume limit, these conical singularities are `blown-up' and smoothed out. The effective 
4-dimensional description then captures this aspect of the geometry by a modulus for each fixed point; the so called blow-up modes.

These blow-up modes can be included in the analysis in a straightforward fashion through the use
of the K\"ahler potential $K$ and superpotential $W$.
Let us give an explicit example; the DGKT model. Here there are 9 fixed points, and so there are 9 blow-up modes. We call these: $\psi_i=a_i+i\,b_i$ for $i=5,\ldots,13$. The expression for the volume in eq.~(\ref{VolDGKT}) is modified to
\begin{equation}
\mbox{Vol} = \mokg_1\mokg_2\mokg_3-\frac{1}{54}\sum_{i=5}^{13}b_i^3.
\end{equation}
The K\"ahler potential (\ref{DGKTKahler}) and the superpotential (\ref{DGKTsuper}) are modified to
\begin{eqnarray}
K &\!\! = &\!\! -\log\left[32\left(\mokg_1\mokg_2\mokg_3-\frac{1}{54}\sum_{i=5}^{13}b_i^3\right)\mokg_4^4\right],\\
W &\!\! = &\!\! \frac{\fs}{\sqrt{2}}  + \sum_{i=1}^3 \frac{\ffi}{\sqrt{2}}\,\moki +\sum_{i=5}^{13}\frac{\ffi}{\sqrt{2}}\,\moki\nonumber\\
&\!\! - &\!\! \frac{\fz}{\sqrt{2}}\left(\moko\,\mokt\,\mokth
-\frac{1}{54}\sum_{i=5}^{13}\moki^3\right) - 2\,\fth\,\moa.\,\,\,\,
\end{eqnarray}
In principle we could now explore this larger moduli space for inflation. However, it is numerically difficult; we have moved from 4 complex moduli (axio-dilation plus three K\"ahler moduli) to 13 complex moduli through the addition of 9 complex blow-up modes.
Instead of a full investigation into the effects of dynamical blow-up modes, we shall freeze the blow-up modes at some vacuum expectation value (vev). Such a vev is explicitly found in the DGKT paper.
We then explore the effect of a non-zero vev for the blow-up modes on the original moduli.
Suppose the $b_i$ are frozen in at some value, which we characterize as: 
$B\equiv\frac{1}{54}\sum_{i=5}^{13}b_i^3$. With $B$ taken as a constant, our K\"ahler potential becomes:
\begin{equation}
K = -\log\left[32\left(\mokg_1\mokg_2\mokg_3 - B\right)\mokg_4^4\right].
\end{equation}

Also, since we are treating the blow-up modes as constants, the superpotential is, for all intents and purposes, unchanged from its value in eq.~(\ref{DGKTsuper}).
This is because we can always shift the real part (axion) of $\moa$ to eliminate any constants.
We perform the same scalings as before in eqns.~(\ref{transfm}), with an extra shift on $\moa$ to eliminate any constants, giving eq.~(\ref{DGKTSuper}). 
Under such field redefinitions we introduce $\bar{B}$, defined such that:
$(\mokg_1\mokg_2\mokg_3-B)\to(\mokg_1\mokg_2\mokg_3-\bar{B})$.

Let us focus on the case in which the axions are vanishing and $\mokg_1=\mokg_2=\mokg_3$,
leaving 2 moduli: $\mokg_1$ and $\moag$. We find the potential:
\begin{eqnarray}
V &\! = &\! V_{\mbox{\tiny{flux}}}
\Big{[}
\left(6\,b_1^2+4\,b_4^2+2\,b_1^6-8 \sqrt{2}\,b_1^3\,b_4\right) \nonumber\\
&\! + &\! \bar{B}\left(12 \sqrt{2}\,b_4 -6\,b_1^3 +\left(4 \delta_a-6\right)/b_1+4 \sqrt{2}\,\delta_b\,b_4/b_1^2\right)\nonumber\\
&\! + &\! \bar{B}^2\left(6+4\,\delta_b/b_1^2+6/b_1^4\right)
 \Big{]}/[32(b_1^3-\bar{B})b_4^4],
\label{DGKTblowup}\end{eqnarray}
where $\delta_a\equiv\hat{f}_1\hat{f}_2+\hat{f}_2\hat{f}_3+\hat{f}_3\hat{f}_1$, 
$\delta_{b}\equiv \hat{f}_0(\hat{f}_1+\hat{f}_2+\hat{f}_3)$, and $V_{\mbox{\tiny{flux}}}$ is given
in eq.~(\ref{DGKTVflux}). There are physical constraints:  $0<\bar{B}<b_1^3$.
Note that the presence of the $\bar{B}$ parameter breaks the scaling
of the model that occurs in the absence of blow-up modes, i.e., scaling only occurs
in the $\bar{B}\to 0$ limit. 
This is because the nonvanishing blow-up modes introduce a non-trivial intersection form, as we encountered previously in the IW model, which prevents the flux parameters from being scaled out completely.
However, several flux parameters can still be eliminated for finite $\bar{B}$.
Note that in the limit $\bar{B}\to 0$ this potential gives precisely the potential in eq.~(\ref{DGKTscaled})
with $\moka_1=\moaa=0$.

Our numerical investigation into $V$ of eq.~(\ref{DGKTblowup}) has again yielded no inflating region,
despite the presence of the tunable parameter $\bar{B}$.

\section{Discussion and Conclusions}\label{Discussion}

We have presented an explicit investigation into three explicit string models.
Although this represents only a rather small part of the landscape, this acts as a starting point
for further investigation into moduli driven inflation.
The non-string theorist should note that despite the inherent complexity of string theory, M-theory etc,
it is possible to strip down the physics in 4 dimensions to familiar territory.
eq.~(\ref{actionK}) gives a familiar 4-dimensional action for $n$ scalar fields minimally coupled
to gravity. We note, however, that the kinetic energy is in general non-canonical since
$K_{i\bar{j}}$ is typically not equal to $\delta_{i\bar{j}}$ and furthermore the geometric moduli and axions appear in the action differently.\footnote{If we ignore the axions and focus on a certain class of simple models, we can perform field redefinitions to obtain a canonical action, as given in 
eq.~(\ref{actionSimple}).} We have proceeded in the usual fashion to check for inflation by examining the slow-roll conditions (\ref{SReps})--(\ref{SReta}).

We have not found inflation in any of the specific models presented.
In the absence of blow-up modes, the DGKT, VZ and IW models involved 8, 14 and 12 real-valued inflaton fields, respectively, making a full numerical exploration of the inflaton 
potential $V(\boldsymbol{\phi})$ 
(which can also be flux dependent) computationally challenging.
We have therefore performed as comprehensive a search as feasible given our available resources:
\begin{itemize}
\item For the DGKT model we derived an analytic expression for the 8-dimensional potential 
$V(\boldsymbol{\phi})$, finding $V(\boldsymbol{\phi})$ to be flux independent (up to an overall scale), and searched the 8-dimensional moduli space for vacua, finding one AdS vacuum in addition to the known SUSY vacuum from \cite{DeWolfe}.
\item For the VZ model, we derived an analytic expression for the 14-dimensional potential 
$V(\boldsymbol{\phi})$, finding that $V(\boldsymbol{\phi})$ depended on the fluxes via a single parameter $s$ (up to an overall scale).
We found the potential to be invariant under permutation of two triplets of complex moduli, and searched the full 6-dimensional subspace corresponding to 
$\psi_1=\psi_2=\psi_3$ , $\psi_5=\psi_6=\psi_7$ for vacua for the cases $s\in\{0,1/2,1,2,5,\infty\}$, finding three new AdS vacua in addition to the known SUSY vacuum from \cite{Villadoro}.
\item For the IW model, we derived an analytic expression for the 12-dimensional potential 
$V(\boldsymbol{\phi})$, finding that $V(\boldsymbol{\phi})$ depended on the fluxes via a single parameter $t$ (up to an overall scale).
We searched the 6-dimensional subspace corresponding to vanishing axions, finding no new vacua in addition to the five AdS vacua reported by \cite{Ihl} for various flux sign combinations.
\end{itemize}
We performed this search for vacua using the numerical packages Mathematica\footnote{http://www.wolfram.com} and 
Singular\footnote{http://www.singular.uni-kl.de} to algebraically solve (using Gr\"obner bases) the set of high order coupled polynomial equations that follow from setting $\nabla V=0$.

We then performed a numerical investigation of the slow-roll conditions, evaluating slow roll parameters for the three models on a multi-dimensional grid 
(of dimension 8, 6, and 6,\footnote{In fact we did a little more than this: In the IW model we did not {\em fully} include the axions, which would be a 12 dimensional space, but did so {\em partially}.} respectively), involving of order $10^{9}$ grid points each,  
and found $\epsilon>1$ for all grid points where $V>0$.
Although we cannot claim to have a {\em proof} of the non-existence of inflation in these
models, as the moduli space is rather large and the potentials $V$ are rather complicated, we do suspect this to be true. We also performed a partial investigation into the consequences of (frozen) blow-up modes, as described in Section \ref{bu}, again finding no inflation.

In the type of models presented we have identified at least three obstacles to realizing inflation: 
the vacua are AdS, there is a logarithmic K\"ahler potential $K$, and suitable field
redefinitions allow one to scale many of the fluxes out of the potential.
None of these features forbid a realization of inflation.
Each is probably a reflection of the simple starting point we took, studying models
closely based on toroidal compactification and focussing on the moduli of the underlying
torus.  
It is certainly known that each of these three points may be avoided in other
regions of the landscape. 
Nevertheless, our result does underscore that slow-roll inflation may be a rare
and delicate phenomenon in the landscape.
We will now discuss each of the three obstacles in turn.

\subsection{The potential energy challenge}

As we discussed earlier in Section \ref{deSitter} it is 
somewhat difficult to realize de Sitter vacua with $V>0$ in string theory. If we break supersymmetry, then existing
analyses suggest that such 
vacua are rare, but plentiful in absolute number. Let us truncate our discussion here to supersymmetric vacua, which 
we know must not 
be de Sitter. A good starting point would be Minkowki vacua which are allowed.
Again focusing on toroidal orientifolds in type IIA string theory, a detailed investigation is
given in \cite{Aldazabal}, in which a host of fluxes are included. In addition to the geometric
fluxes that we have described, they turn on so-called non-geometric fluxes \cite{Jessie},
and additionally turn on fluxes associated with S-duality (strong -- weak coupling duality).
In this framework, although they are non-generic, Minkowski vacua are explicitly found (see
also \cite{Vafa}). 
This may be a good starting point for considering inflation models where the inflaton eventually
settles down to zero energy density. However, the Minkowski vacua given in \cite{Aldazabal}
are not under good perturbative control. In other words, it is expected that there
are large $\alpha'$ and $g_s$ (loop) corrections to the potential. This is in contrast to the models
we have investigated in this article. In each case we could dial the fluxes in a particular
fashion so that all quantum corrections were small. This justifies the supergravity treatment
and makes the results of our investigation particularly informative.

\subsection{The kinetic energy challenge}

Let us turn to the form of the kinetic energy, which is governed by the K\"ahler potential $K$.
As we have pointed out several times, in supergravity models this is typically logarithmic.
The Hessian matrix of $K$ determines the form of the kinetic terms. The tree-level form of this for
the diagonal torus model is given in eq.~(\ref{actioncan}). At the level of supergravity (i.e., ignoring quantum
corrections) this form is rather generic for non-torus models also \cite{Candelas}. 
So for instance, this kind of metric on moduli space will generically occur for the volume modulus
$\gmog.$
Let us write this form as:
\begin{equation}
T\sim-\partial_\mu(\log\gmog)\partial^\mu(\log\gmog)
\label{Tlog}\end{equation}
(suppressing factors of $\mpl$).
Although we shall not go through the explicit details here, this is fairly simple
to show from the fact that in performing the dimensional reduction from 10 to 4 dimensions,
we pick up factors of the volume modulus.
In order to move to the Einstein frame, we must then compute the transformation of the Ricci scalar, which is a contraction of the Riemann tensor. Since the mixed derivative terms of the Riemann tensor
are proportional to the Christoffel symbols, and since the Christoffel symbols essentially perform
{\em logarithmic} derivatives of the metric, the result in (\ref{Tlog}) follows.

For models involving fluxes etc., it is rather generic that the potential $V$ be given
(at large volume) by
some rational function of $b$ (or more generally of the full set of K\"ahler moduli). 
We have given several explicit examples of this in this paper.
Let us drastically simplify the form by writing
\begin{equation}
V\sim \sum_i ~c_i\,\gmog^{-k_i}
\label{Vvolmod}\end{equation}
where the $k_i$ are positive integers and the $c_i$ are some coefficients.\footnote{Locality
in the extra dimensions allows one to prove that any contribution to the Einstein-frame
potential should fall at least as
quickly as $1/r^6$ at large radius $r$ for the extra dimensions (as described on e.g.
page 12 of \cite{KMS}).  This puts a bound on 
the $\vert k_i \vert$, and explains why we have disallowed contributions which grow at large
radius.}
We note that by defining $\phi\equiv \log\gmog$, the kinetic
energy is in canonical form and $V\sim \sum_i~c_i e^{- k_i \phi}$. 
(This property was alluded to earlier in eq.~(\ref{actionSimple})).
 Of course exponentials have slow-roll parameters $\epsilon\sim\eta\sim k^2$, which are not generically
small for positive integers $k$. A recent discussion of how one can perhaps
construct working models by fine-tuning
similar potentials with several terms appears in \cite{Itzhaki}. We point out that in \cite{Itzhaki},
only the volume modulus is considered and other moduli are treated as fixed; 
we have seen explicit examples in our models where although one partial derivative of the potential may be small,
another one will often be large, hence making $\epsilon$ large and spoiling inflation.
 
\subsubsection{$\alpha'$ corrections}
 
Let us comment now on the effect that $\alpha'$ corrections have. According to 
\cite{Becker}, in type IIB string theory there exists an $\alpha'$ correction to the K\"ahler potential 
from an $\mathcal{O}\!\left(R^4\right)$ term in the 10-dimensional action. 
In units where $2\pi\alpha'=1$, the piece coming from the volume modulus is found to be
\begin{equation}
K=-2\ln\left[(2 \gmog)^{3/2}+\hat{\xi}\right]
\end{equation}
where $\hat{\xi}=-\zeta(3)\chi e^{-3\phi/2}/4$ with $\chi$ the Euler characteristic of the Calabi-Yau. The dilaton $\phi$, and hence $\hat{\xi}$, is assumed to be fixed.
Although we are considering IIA orientifolds (which are not directly related to IIB models, because of the 
flux), let us imagine the effects of a similar correction in our context.
Our models have $\chi$ which is of ${\cal O}(1-100)$ -- although the torus has vanishing $\chi$, the 
fixed points of the orbifold group action introduce blow-up modes that generate non-trivial $\chi$.
However, in the regime where our classical analysis is trustworthy (and by choosing sufficiently large
fluxes we can make it arbitrarily reliable \cite{DeWolfe}), we can neglect this effect.
One could imagine that for more general Calabi-Yau's $\chi$, and hence the $\hat{\xi}$ correction, 
is sometimes large. 

It is known that such a term can indeed be important for inflation, see e.g. \cite{Weltman}.
For $\gmog^{3/2}\gg\hat{\xi}$, this correction is irrelevant and we return to the previous analysis.
For $\gmog^{3/2}\ll\hat{\xi}$, however, this changes the situation considerably. In this case, one finds that the kinetic term for $\gmog$ is modified to, roughly
\begin{equation}
T\sim-\frac{1}{\hat{\xi}\sqrt{\gmog}}(\partial_\mu \gmog)^2.
\end{equation}
So the inverse of the metric on moduli space is $K^{i\bar{j}}\sim\hat{\xi}\sqrt{b}$.
The inclusion of $\alpha'$ corrections into the K\"ahler potential also induces corrections to the potential $V$ through the supergravity formula. However, let us
again assume the form for $V$ as given in eq.~(\ref{Vvolmod}) for the purposes of illustration. We find that generically
$\epsilon\sim\eta\sim k^2\,\hat{\xi}/b^{3/2}$, which is large in the assumed regime $\gmog^{3/2}\ll\hat{\xi}$.
Ref.~\cite{Weltman} shows that even in the 
setups containing non-perturbative corrections, e.g., race-track etc, this $\alpha'$
correction usually makes achieving inflation harder or impossible.
Of course this entire discussion should be viewed with caution: in the regime where the $\hat{\xi}$-correction
has a significant effect, one would have to carefully justify any analysis which neglects additional
$\alpha^\prime$ 
corrections.  

\subsubsection{Approximate K\"ahler potentials}

One further comment on the form of the kinetic energy comes from ``inflation in supergravity" treatments,
e.g., \cite{Kawasaki}.
There it is often assumed that the K\"ahler potential takes on the minimal form: $K=\phi^*\phi$ 
(giving $K_{i\bar{j}}=\delta_{i\bar{j}}$). 
This form does not literally occur in any string compactifications that we are aware of. 
It can appear as an approximate K\"ahler potential in models where one fixes the moduli of the compactification manifold
and expands the K\"ahler potential for brane position moduli (or, sometimes, axions) in a Taylor expansion.
So we believe that realizing inflation in these scenarios should be taken with a grain of salt, subject to justifying
the appearance of the relevant $K$, for the relevant range of field space, in a model with fixed moduli.

\subsection{The challenge of fluxes scaling out}

Turning to the issue of scaling out fluxes, although this occurs in the DGKT model if one neglects blow-up
modes and focuses on untwisted moduli, 
it did not occur in full in the other VZ and IW models. In the DGKT model, neglecting the blow-up modes, every member of
the infinite set of vacua was identical from the point of view of the slow-roll conditions,
and there was large, but not complete, degeneracy in the other models.
Degeneracy is reduced in the presence of blow-up modes.
In general though, the ability to exploit scaling and shift symmetries 
reduces the freedom 
allowed in dialing the shape of a potential in the landscape.  Much like the relation whereby unbroken supersymmetry
generically implies AdS vacua, or the simple geometric arguments which determine the logarithmic
form of the K\"ahler potential, this all points to the idea that the landscape,
although extremely large, has {\em structure}.
On the other hand, the relatively simple form of flux potentials for toroidal moduli,
which is behind the existence of some of these scaling symmetries, would not persist in generic Calabi-Yau
models.  Therefore, it is reasonable to postulate that the
degeneracy we found may be an artifact of the particular simple models we have examined.

\subsection{Outlook}

Alternatively, it could be that the type of construction discussed in the introduction,
where inflation is realized through a combination of ingredients, including non-perturbative
corrections to the superpotential, is more promising.  For example, let us make a comment on
the $N$-flation idea \cite{Nflation_1,Nflation_2}, 
which requires $N$ massless axions at the perturbative level,
whose mass is then generated by non-perturbative effects.
We have seen one flat direction of the axions
in the untwisted modes of the VZ model, and the papers \cite{DeWolfe,Nflation_1,Nflation_2} discuss how one can have $N\gg 1$ for
more complicated compact spaces. 
However, various model building assumptions made in \cite{Nflation_1,Nflation_2} can certainly be questioned, and
an explicit realization of this class of scenarios is important to unravel.

In summary, our work should be viewed as a starting point for  
a much more general study into inflation driven by computable flux potentials. 
One obvious next step would be to study a similar class of problems in more general Calabi-Yau
manifolds, rather than orbifolds of the torus. 
The more complicated structure of the internal geometry should translate into
richer flux potentials, which could solve some of the problems we found in the toroidal models.
Another approach could be to develop statistical arguments along the lines of \cite{DD} to
quantify how generically or non-generically one expects to find inflation in flux vacua.

\section*{Acknowledgments}

We would like to thank Liam McAllister and Washington Taylor for numerous helpful discussions and
for detailed comments on drafts of this paper.
We would also like to thank
Rizwan Ansari, Andrea De Simone, Thomas Faulkner, Thomas Grimm, Renata Kallosh, Abhishek Kumar, 
Vijay Kumar, Hong Liu, Courtney Peterson, Carlos Santana, Eva Silverstein, Molly Swanson, David Vegh, Alexander Vilenkin, Giovanni Villadoro, Frank Wilczek, and Timm Wrase for helpful discussions.
This work was supported by NASA grants NAG5-11099 and NNG06GC55G,
NSF grants AST-0134999, AST-0607597, AST-0708534 and PHY-0244728,
the U.S. Department of Energy under contracts DE-AC03-76SF00515 and DE-FC02-94ER40818, 
the Kavli Foundation, and fellowships
from the David and Lucile Packard Foundation and the Research Corporation.


\clearpage
\appendix

\section{DGKT Potential}
In the DGKT setup, there are 4 complex pairs of moduli: $\mokao,\,\mokgo,\ldots,\moaa,\,\moag$.
Using the K\"ahler potential of eq.~(\ref{DGKTKahler}) and the superpotential of eq.~(\ref{DGKTSuper}) we find the 
following 4-dimensional potential:
\begin{widetext}
\begin{eqnarray}
V=V_{\mbox{\tiny{flux}}}\big{[} \amp \amp \!\!\!\!\!
2(\mokao + \mokat + \mokath + 2\,\sqrt{2}\,\moaa)^2
- 4\,\sgn\,\mokao\,\mokat\,\mokath (\mokao+\mokat+\mokath+2\,\sqrt{2}\,\moaa) + 2\,\mokao^2\,\mokat^2\,\mokath^2   
\nonumber\\
\amp + \amp 2(\mokgo^2+\mokgt^2+\mokgth^2)+ 4\,\moag^2 - 4\,\sgn\,(\mokat\,\mokath\,\mokgo^2 + \mokao\,\mokath\,\mokgt^2
  + \mokao\,\mokat\,\mokgth^2) \nonumber \\
\amp+\amp 2(\mokat^2\,\mokath^2\,\mokgo^2 + \mokao^2\,\mokath^2\,\mokgt^2 + \mokao^2\,\mokat^2\,\mokgth^2)
+ 2(\mokath^2\,\mokgo^2\,\mokgt^2 + \mokat^2\,\mokgo^2\,\mokgth^2 + \mokao^2\,\mokgt^2\,\mokgth^2) 
  + 2\,\mokgo^2\,\mokgt^2\,\mokgth^2 \nonumber \\
\amp-\amp 8\,{\sqrt{2}}\,\mokgo\,\mokgt\,\mokgth\,\moag  \big{]}/(32\,\mokgo\,\mokgt\,\mokgth\,\moag^4),
\label{DGKTfull}\end{eqnarray}
\end{widetext}
with $V_{\mbox{\tiny{flux}}}$ given in (\ref{DGKTVflux}) and $\sgn=\hfz\hffo\hfft\hffth=\pm 1$. 
The slow-roll parameter $\epsilon$ is given by (\ref{epsilon}) with $n_1=n_2=n_3=1,\,n_4=4$.

\section{VZ Potential}

In the VZ setup, there are 7 complex pairs of moduli: $\vmokao,\,\vmokgo,\ldots,\vmocath,\,\vmocgth$.
Let us focus on the symmetric case $\vmokao=\vmokat=\vmokath$, $\vmokgo=\vmokgt=\vmokgth$,
$\vmocao=\vmocat=\vmocath$, $\vmocgo=\vmocgt=\vmocgth$.
Using the K\"ahler potential of eq.~(\ref{VZKahler}) and the superpotential of eq.~(\ref{VZSuper}) in the symmetric case, we find the following 4-dimensional potential:
\begin{widetext}
\begin{eqnarray}
V=V_{\mbox{\tiny{flux}}}\!\!\amp[\amp
36 a_1^6-72 \rat  a_1^5+36 \rat ^2 a_1^4+108 b_1^2 a_1^4+144 \rat  \hat{a}_4 a_1^4+216 a_1^4-144 \rat  b_1^2 a_1^3+144 \rat  a_1^3-144 \rat
   ^2 \hat{a}_4 a_1^3\nonumber\\
   \amp - \amp 48 \hat{a}_4 a_1^3+108 b_1^4 a_1^2-360 \rat ^2 a_1^2+144 \rat ^2 \hat{a}_4^2 a_1^2+48 \rat ^2 b_1^2 a_1^2+144 \rat  \hat{a}_4 b_1^2 a_1^2+216
   b_1^2 a_1^2+144 \rat ^2 b_4^2 a_1^2 \nonumber\\
\amp + \amp 432 \rat ^2 b_5^2 a_1^2+480 \rat  \hat{a}_4 a_1^2+324 a_1^2-72 \rat  b_1^4 a_1-96 \rat  \hat{a}_4^2 a_1-144
   \rat  b_1^2 a_1-96 \rat ^2 \hat{a}_4 b_1^2 a_1-96 \rat  b_4^2 a_1 \nonumber\\
\amp - \amp 288 \rat  b_5^2 a_1+1080 \rat  a_1+720 \rat ^2 \hat{a}_4 a_1-144 \hat{a}_4 a_1+36
   b_1^6+12 \rat ^2 b_1^4+900 \rat ^2+16 \hat{a}_4^2+48 \rat ^2 \hat{a}_4^2 b_1^2 \nonumber\\
\amp + \amp 144 \rat  \hat{a}_4 b_1^2+108 b_1^2+48 \rat ^2 b_1^2 b_4^2+16 b_4^2-432
   \rat ^2 b_1^2 b_5^2+48 b_5^2-240 \rat  \hat{a}_4+48 \rat ^2 b_1^3 b_4-48 b_1^3 b_4 \nonumber\\
\amp + \amp 144 \rat ^2 b_1^3 b_5-144 b_1^3 b_5-576 \rat ^2 b_1^2 b_4 b_5 ]/(b_1^3 b_4 b_5^3),
\label{VZfull}\end{eqnarray}
\end{widetext}
with $V_{\mbox{\tiny{flux}}}$ given in (\ref{VZVflux}), $\hat{a}_4=a_4+3 a_5$, and $\rat=\None/\Ntwo$. 
The slow-roll parameter $\epsilon$ is given by (\ref{epsilon}) with $n_1=\ldots=n_7=1$.

\section{IW Potential}
In the IW setup, there are 6 complex pairs of moduli: $\imokao,\,\imokgo,\ldots,\imoca,\,\imocg$.
Using the K\"ahler potential of eq.~(\ref{IWKahler}) and the superpotential of eq.~(\ref{IWSuper}), we find the 
following 4-dimensional potential:
\begin{widetext}
\begin{eqnarray}
V=V_{\mbox{\tiny{flux}}}\amp\!\!\big{[}\amp\!\!
a_3^2 a_4^4/2+b_3^2 a_4^4/2-2 a_1 a_2 a_3^2 a_4^2-2 a_1 a_2 b_3^2 a_4^2+2 b_1 b_2 b_3^2 a_4^2+a_3^2 b_4^2 a_4^2+b_3^2 b_4^2 a_4^2+2 \sgn
    a_2 a_3 a_4^2 \nonumber\\
\amp+\amp 2 a_3^2 b_1 b_2 a_4^2+2\sgnthz a_3^2  a_4^2+2\sgnthz b_3^2 a_4^2+2\sgnthz a_1 a_3 a_4^2+4 \sqrt{2}\sgnthz a_3 a_5 
   a_4^2+4 \sqrt{2}\sgnthz a_3 a_6 a_4^2 \nonumber\\
\amp-\amp 4 a_2 b_1 b_3^2 b_4 a_4-4 a_1 b_2 b_3^2 b_4 a_4-4 a_2 a_3^2 b_1 b_4 a_4-4 a_1 a_3^2 b_2 b_4 a_4+4 \sgn  a_3 b_2
   b_4 a_4+4\sgnthz a_3 b_1 b_4 a_4\nonumber\\
\amp+\amp a_3^2 b_4^4/2+b_3^2 b_4^4/2+2 a_1^2+2 a_2^2+2 a_1^2 a_2^2 a_3^2+2 a_3^2+16 a_5^2+16 a_6^2+2 a_2^2
   a_3^2 b_1^2+2 b_1^2+2 a_1^2 a_3^2 b_2^2\nonumber\\
\amp+\amp 2 a_3^2 b_1^2 b_2^2-4 \sgn  a_1 a_3 b_2^2+2 b_2^2+2 a_1^2 a_2^2 b_3^2+2 a_2^2 b_1^2 b_3^2+2 a_1^2 b_2^2 b_3^2+2 b_1^2
   b_2^2 b_3^2+2 b_3^2+2 a_1 a_2 a_3^2 b_4^2\nonumber\\
\amp+\amp 2 a_1 a_2 b_3^2 b_4^2-2 b_1 b_2 b_3^2 b_4^2-2 \sgn  a_2 a_3 b_4^2-2 a_3^2 b_1 b_2 b_4^2+8 b_5^2+8 b_6^2-4 \sgn 
   a_1 a_2^2 a_3+4 a_1 a_3\nonumber\\
\amp+\amp 8 \sqrt{2} a_1 a_5+8 \sqrt{2} a_3 a_5+8 \sqrt{2} a_1 a_6+8 \sqrt{2} a_3 a_6+32 a_5 a_6+4 \sqrt{2} b_3 b_4^2 b_5-8 \sqrt{2} b_1 b_2 b_3
   b_5\nonumber\\
\amp+\amp 4 \sqrt{2} b_3 b_4^2 b_6-8 \sqrt{2} b_1 b_2 b_3 b_6+2\sgnot b_4^2 +4\sgnot a_1 a_2 +4\sgnot a_2 a_3 +8 \sqrt{2} a_2 a_5 \sgn
   _{12}+8 \sqrt{2}\sgnot a_2 a_6\nonumber\\
\amp-\amp 4\sgnthz a_1 a_2 a_3^2 -4\sgnthz a_2 a_3 b_1^2 -4\sgnthz a_1 a_2 b_3^2 -2\sgnthz a_1 a_3 b_4^2 
 -4\sgnthz a_1^2 a_2 a_3 -8 \sqrt{2}\sgnthz a_1 a_2 a_3 a_5 \nonumber\\
\amp-\amp 8 \sqrt{2}\sgnthz a_1 a_2 a_3 a_6  +(2\sgnthz a_3  a_4^3+4
   a_1 a_4+4 a_3 a_4+8 \sqrt{2} a_5 a_4+8 \sqrt{2} a_6 a_4+4\sgnot a_2 a_4 \nonumber\\
\amp+\amp 2\sgnthz a_3 b_4^2 a_4-4\sgnthz a_1 a_2 a_3 a_4+4\sgnthz a_3 b_1 b_2
   a_4+4 b_1 b_4+4\sgnot b_2 b_4 -4\sgnthz a_2 a_3 b_1 b_4 \nonumber\\
\amp-\amp 4\sgnthz a_1 a_3 b_2 b_4 ) \flx  + (2 a_4^2+b_4^2+2 b_1 b_2) \flx^2
\big{]}/[2\,(\imokgo\imokgt-\imokgf^2/2) \imokgth \imoag^2 \imocg^2],
\label{IWfull}\end{eqnarray}
\end{widetext}
with $V_{\mbox{\tiny{flux}}}$ given in (\ref{IWVflux}), $\flx$ given in (\ref{IWflx}), 
$\sgn=\hfz\hffo\hfft\hffth=\pm 1,\,\sgnot=\hffo\hfft=\pm 1,\,\sgnthz=\hffth\hfz=\pm 1$. 
The slow-roll parameter $\epsilon$ is given by (\ref{IWeps}).

\section{Cosmological Parameters from Slow-Roll Inflation}\label{cookbook}

The mathematical prescription in this Section allows one to compute cosmological parameters corresponding to an arbitrary string potential without understanding the derivation or interpretation of the results. 

Suppose from some string model we are given a potential energy function $\VE$ 
of some complex scalar fields $\psi^i$ in the Einstein frame (see eq.~(\ref{actionK}))
and a K\"ahler potential $K$. For example, $\VE$ may be given by
the supergravity formula in eq.~(\ref{SUGRA}) complemented by the rescaling eq.~(\ref{VEinstein}).
We then compute the following slow-roll parameters:
\begin{eqnarray}
&& \epsilon = \frac{K^{i\bar{j}}\VE_{,i}\VE_{,\bar{j}}}{\VE^2} \left( =\frac{g^{ab}\VE_{,a}\VE_{,b}}{2\VE^2}\right), \label{SReps2}\\
&& \eta = \mbox{min eigenvalue}\left\{\frac{g^{ab}\left(\VE_{,bc}-\Gamma^d_{bc}\VE_{,d}\right)}{\VE}\right\},
\label{SReta2}
\end{eqnarray}
where $\eta$ (and $\epsilon$) is written in terms of the metric $g_{ab}$ governing real scalar fields $\phi^a$:
$K_{i\bar{j}}\partial_\mu\psi^i\partial^\mu\psi^{\bar{j}}=\frac{1}{2}g_{ab}\partial_\mu\phi^a\partial^\mu\phi^b$, with $\phi^{2i-1}=$ Re$[\psi^i]$ and $\phi^{2i}=$ Im$[\psi^i]$.

The universe inflates until a time $t_e$ when the slow-roll conditions ($\epsilon<1$, $|\eta|<1$) 
are no longer satisfied. The number of $e$-foldings from time $t$ to $t_e$ is defined by
\begin{equation}
N=\int_{t}^{t_e} dt\,H .
\end{equation}
All the cosmological parameters defined below are a function of $N$. 
A good value to use is 55 (see \cite{Liddle}), with a reasonable range being $50<N<60$.

According to inflation, several cosmological parameters can be computed as follows:
\beqa{dHeq}
Q_s& = &\sqrt{\frac{\VE}{150\pi^2\mpl^4\epsilon}},\\
\ns& = &1-\partial_N\ln Q_s^2=1-6\epsilon+2\eta,\\
\alpha_s& = &-\partial_N^2\ln Q_s^2,\\
Q_t& = &\sqrt{\VE\over 75\pi^2\mpl^4},\,\,\, r\equiv\left({Q_t\over Q_s}\right)^2=16\epsilon,\\
\nt& = & -\partial_N Q_t^2 = -2\epsilon,
\eeqa
which corresponds to the amplitude, spectral index, and running of spectral index of scalar fluctuations, and the amplitude and spectral index of tensor fluctuations, respectively. 
The expressions giving $Q_t$ and $n_t$ have general validity. In contrast, the expressions for $Q_s$, $n_s$ and $\alpha_s$ 
are good approximations for the most studied cases of multi-field inflation in the literature, where 
the walls of the multi-dimensional gorge in which the inflaton slowly rolls are much steeper than the 
roll direction, but do not hold more generally. The expression for $Q_s$ always provides a lower limit on the correct value.

The predictions for these cosmological parameters can be directly compared with with observation. 
The most recent constraints from combining WMAP (Wilkinson Microwave Anisotropy Probe) microwave background data with SDSS (Sloan Digital Sky Survey) galaxy clustering data are \cite{LSS}
\beqa{MeasurementEq}
Q_s&=&1.945^{+ 0.051}_{- 0.053}\times10^{-5},\\
\ns&=&0.953^{+ 0.016}_{- 0.016},\\
\alpha_s&=&-0.040^{+ 0.027}_{- 0.027},\\
r&<&0.30 \>(95\%),\\
\nt + 1 &=&0.9861^{+ 0.0096}_{- 0.0142}.
\eeqa

\end{document}